\documentclass[a4paper,11pt]{article}
\pdfoutput=1 

\usepackage{jcappub} 

\usepackage[T1]{fontenc} 
\usepackage[normalem]{ulem}

\title{Constraining UV freeze-in of light relics with current and next-generation CMB observations}

\author[a,b]{Luca Caloni,}
\author[b]{Patrick Stengel,}
\author[b]{\\Massimiliano Lattanzi,}
\author[b]{Martina Gerbino}

\affiliation[a]{Dipartimento di Fisica e Scienze della Terra, Universit\`{a} degli Studi di Ferrara,\\Via Giuseppe Saragat 1, I-44122 Ferrara, Italy}
\affiliation[b]{Istituto Nazionale di Fisica Nucleare, Sezione di Ferrara,\\Via Giuseppe Saragat 1, I-44122 Ferrara, Italy}

\emailAdd{luca.caloni@unife.it}
\emailAdd{pstengel@fe.infn.it}
\emailAdd{lattanzi@fe.infn.it}
\emailAdd{gerbinom@fe.infn.it}

\abstract{Cosmological observations allow to measure the abundance of light relics produced in the early Universe. Most studies focus on the thermal freeze-out scenario, yet light relics produced by freeze-in are generic for models in which new light degrees of freedom do not couple strongly enough to the Standard Model (SM) plasma to allow for full thermalization in the early Universe. In ultraviolet (UV) freeze-in scenarios, rates for light relic production associated with non-renormalizable interactions typical of beyond the SM (BSM) models  grow with temperature more quickly than the Hubble rate. Thus, relatively small couplings to the SM can be probed by current and next-generation cosmic microwave background (CMB) experiments. We investigate several representative benchmark BSM models, such as axion-like particles from Primakoff production, massless dark photons and light right-handed neutrinos. We calculate contributions to the effective number of neutrino species, $\Delta N_{\rm eff}$, in corners of parameter space not previously considered and discuss the sensitivity of CMB experiments compared to other probes. In contrast to freeze-out scenarios, $\Delta N_{\rm eff}$ from UV freeze-in is more dependent on both the specific BSM physics model and the reheating temperature. Depending on the details of the BSM scenario, we find that the sensitivity of next-generation CMB experiments can complement or surpass the current astrophysical, laboratory or collider constraints on the couplings of the SM to the light relic.}

\begin{document}
\maketitle
\flushbottom

\section{Introduction}
\label{sec:intro}

The production of light relics in the early Universe is predicted by many extensions of the Standard Model (SM) and can be probed by current and next-generation measurements of the cosmic microwave background (CMB). Axion-like particles (ALPs) are pseudo Nambu-Goldstone bosons~\cite{PhysRevLett.40.223,PhysRevLett.40.279,PhysRevLett.40.220} which can arise from the spontaneous breaking of global symmetries in scenarios beyond the SM (BSM)~\cite{McDonald:1993ex,Burgess:2000yq,He:2009yd,Cacciapaglia:2019bqz} and are characteristic of string theory models with compactified extra dimensions~\cite{Svrcek:2006yi,Arvanitaki:2009fg}. Gauged $U(1)$ symmetries under which no SM particles are charged can yield dark photons (DPs)~\cite{Holdom:1985ag,Fabbrichesi:2020wbt}. While typical in certain string theory compactifications~\cite{Abel:2003ue,Abel:2006qt,Abel:2008ai,Goodsell:2009xc}, DPs can also be present in mirror world scenarios~\cite{Berezhiani:2003xm,Berezhiani:2008gi,Salvio:2014soa} and the low-energy spectrum of asymptotically free field theories~\cite{Salvio:2020prd,Ghoshal:2020vud}. Light right-handed neutrinos (RH$\nu$s) can be present in see-saw models which explain the smallness of active neutrino masses~\cite{Mohapatra:2005wg,King:2013eh,King:2015aea} and are necessary for anomaly cancellation in scenarios with gauge symmetries beyond those in the SM~\cite{Carlson:1986cu,Feldman:2011ms}.    

Light relics can impact a variety of cosmological observables at CMB experiments, including the effective number of neutrino species, $N_{\rm eff}$~\cite{Dvorkin:2022jyg,Bashinsky:2003tk,Hou:2011ec,Baumann:2015rya}. The imprint of additional light species on cosmology depends on the mechanism by which the particles are produced. Both thermal and non-thermal production mechanisms can potentially yield observational signatures at CMB experiments, with the former typically considered to be less model dependent so long as the new light species are in thermal equilibrium with the SM plasma at some point in the early universe. For light species in thermal equilibrium, contributions to $N_{\rm eff}$ can be parameterized simply by the strength of the relevant couplings to the SM or, equivalently, the temperature at which the species decouple from the thermal bath. These freeze-out scenarios have been well-explored in the literature generically for light species with different spins, and applied to a wide variety of BSM physics paradigms~\cite{Brust:2013ova,Chacko:2015noa,Baumann:2016wac}.

Freeze-in scenarios in which light relics are produced through interactions with the thermal bath but never are in thermal equilibrium, while more dependent on both the specific BSM physics model and the reheating temperature, are also rather generic for models in which the light relic couples to the SM plasma more weakly than necessary for full thermalization~\cite{Hall:2009bx,Cheung:2010gj,Hall:2010jx,Elahi:2014fsa,DEramo:2020gpr}. Depending on the type of interactions, freeze-in production typically occurs predominantly at temperatures either close to or much higher than the masses of the SM particles coupled to the light relic. Usually referred to as infrared (IR) freeze-in~\cite{Hall:2009bx,Cheung:2010gj,Hall:2010jx,DEramo:2020gpr}, the former case is characterized by renormalizable interactions between the light relic and the SM which become most efficient just before the equilibrium densities of the thermalized SM species become Boltzmann suppressed. Alternatively, scenarios with non-renormalizable SM interactions tend to produce light relics most efficiently at the highest possible temperature at which the SM species are thermalized, often the reheating temperature, and are thus referred to as ultraviolet (UV) freeze-in~\cite{Hall:2009bx,Elahi:2014fsa}.

In this work, we focus on the deviation from the expected SM value for the effective number of relativistic species, $N_{\rm eff}^{\rm SM} = 3.044$ \cite{Mangano:2001iu,Bennett:2019ewm,Bennett:2020zkv,Akita:2020szl,Froustey:2020mcq,Cielo:2023bqp,Drewes:2024wbw}, associated with UV freeze-in production of massless ALPs, DPs and RH$\nu$s. The strongest bound on $N_{\rm eff}$ comes from Planck$+$BAO data, which yield $N_{\rm eff} = 2.99^{+0.34}_{-0.33}$ at 95\% CL \cite{Planck:2018vyg}. For a positive deviation from $N_{\rm eff}^{\rm SM}$, Planck$+$BAO data implies the upper limit $\Delta N_{\rm eff} < 0.30$ at 95\% CL \cite{Planck:2018vyg}. Future CMB surveys, like the Simons Observatory (SO) \cite{SimonsObservatory:2018koc} and CMB-S4 \cite{CMB-S4:2016ple}, will measure $\Delta N_{\rm eff}$ with an improved precision, with a forecasted sensitivity of $\sigma(\Delta N_{\rm eff})_{\rm SO} = 0.05$ and $\sigma(\Delta N_{\rm eff})_{\rm{CMB-}S4} = 0.03$, respectively. The proposed CMB-HD survey \cite{CMB-HD:2022bsz} would further improve these measurements, reaching a sensitivity of $\sigma(\Delta N_{\rm eff})_{\rm{CMB-}HD} = 0.014$. By tracking the dynamics of light relic freeze-in using the Boltzmann equation relevant for the evolution of the number density in each scenario, we demonstrate how current and next-generation CMB observations are sensitive to parameter space in which the light relics were never in thermal equilibrium. 

Our analysis thus extends and complements previous work considering cosmological constraints on light relics. Recent work considering cosmological constraints on thermal ALP production has typically focused on the specific case of the QCD axion~\cite{DEramo:2022nvb,Notari:2022ffe,Bianchini:2023ubu} or thermal production of ALPs in equilibrium~\cite{Dror:2021nyr,Caloni:2022uya}. A detailed calculation of ALP freeze-in has been considered in Ref.~\cite{DEramo:2023nzt} for scattering production initiated by generic (massive) particles in the thermal bath. Alternatively, we investigate the concrete example of Primakoff production~\cite{Bolz:2000fu, Cadamuro:2011fd} and broadly explore the parameter space for UV freeze-in of ALPs associated with an ALP-photon coupling. Similarly, most previous studies of massless DPs have assumed thermalization with the SM plasma~\cite{Dobrescu:2004wz,Vogel:2013raa,Foot:2014uba,Salvio:2022hfa}. Ref.~\cite{Adshead:2022ovo} considered the production of DPs through renormalizable interactions with the SM without assuming thermal equilibrium. Our work is thus a complementary investigation of how CMB observations can constrain UV freeze-in of DPs through non-renormalizable interactions.   

We investigate the freeze-in production of RH$\nu$s for extensions of the SM in which baryon minus lepton number ($B-L$), a global symmetry within the SM, is promoted to a gauge symmetry. This $U(1)_{B-L}$ results in a new (massive) gauge boson $Z'$ and RH$\nu$s are required for the cancellation of gauge anomalies. We only consider scenarios in which the mass of the $Z'$ is significantly higher than the reheat temperature, such that UV freeze-in of RH$\nu$s proceeds through effective contact interactions with SM fermions. As for ALPs and DPs, UV freeze-in production of RH$\nu$s has typically been previously studied in the parameter space relevant for thermalization with the SM~\cite{Barger:2003zh,Heeck:2014zfa,FileviezPerez:2019cyn,Abazajian:2019oqj}. Ref.~\cite{Adshead:2022ovo} also investigated the production of RH$\nu$s in gauged $B-L$ models without assuming thermal equilibrium for reheat temperatures much larger than the mass of the $Z'$, again complementary to the parameter space considered in this study. 

In addition to gauged $B-L$ models, we also consider the possibility of RH$\nu$s produced by interactions induced by the neutrino charge radius~\cite{Bernabeu:2004jr}. Constraining the electromagnetic properties of neutrinos using cosmological observations has been the subject of previous work, with a major focus on the neutrino magnetic moment. Production of light RH$\nu$s through the neutrino magnetic moment was studied in \cite{Elmfors:1997tt} and recently reassessed in \cite{Li:2022dkc}, both assuming thermal freeze-out. Ref. \cite{Carenza:2022ngg} derives constraints on the neutrino magnetic moment for freeze-in as well as freeze-out production, with an approach similar to the one used in this paper. Here, we focus on light RH$\nu$s produced by fermion-antifermion annihilations induced by the neutrino charge radius. Our analysis extends previous works~\cite{Grifols:1986ed, Grifols:1989vi} by considering RH$\nu$ production not in equilibrium with the SM.

The rest of this paper is structured as follows. In Sec.~\ref{sec:UV-freezein}, we describe our general approach to solving the relevant Boltzmann equations and calculating contributions to $\Delta N_{\rm eff}$ for both freeze-out and UV freeze-in scenarios. In Sec.~\ref{sec:axions}, we calculate upper limits on the ALP-photon coupling due to the Primakoff effect and compare to laboratory and astrophysical constraints. In Sec.~\ref{sec:dark-photons}, we calculate constraints on UV freeze-in production of DPs, for scenarios most relevant to reheat temperatures either above or below the electroweak phase transition. In Sec.~\ref{sec:sterile-neutrinos-B-L}, we calculate upper limits on the $U(1)_{B-L}$ gauge coupling from UV freeze-in production of RH$\nu$s and compare to collider constraints. In Sec.~\ref{sec:chargeradius}, we calculate upper limits on the charge radius of RH$\nu$s and compare to astrophysical constraints. We conclude with a brief summary of our results and a discussion of the outlook for future work in Sec.~\ref{sec:conclusions}.

\section{Boltzmann equation and \texorpdfstring{$\Delta N_{\rm eff}$}{} from light relics}	
\label{sec:UV-freezein}
Let us consider a light particle species $\phi$, which is produced in the early Universe with a rate $\Gamma_\phi$. 
During the radiation-dominated era, the Universe expands at a rate
\begin{equation}
    H(T) = \frac{\pi}{\sqrt{90}}g_*(T)^{1/2}\frac{T^2}{M_{\rm Pl}} \, ,
\end{equation}
where $M_{\rm Pl} = (8\pi G)^{-1/2} \simeq 2.43 \times 10^{18} \; {\rm GeV}$ is the reduced Planck mass and $g_*(T)$ is the effective number of relativistic degrees of freedom.
The production mechanisms considered in this paper have in the final state a number of particles of the species $\phi$ given by $l = 1$ (e.g., Primakoff effect for ALPs) or $2$ (e.g., fermion-antifermion annihilations for RH$\nu$s).
The comoving number density of $\phi$, $Y_\phi \equiv n_\phi/s$, evolves according to the following Boltzmann equation
\begin{equation}
	\label{eq:Boltzmann}
	\frac{dY_\phi}{d\log x} - \left(1 - \frac{1}{3}\frac{d\log g_{*s}}{d\log x}\right) \frac{\Gamma_\phi(x)}{H(x)} \left[Y_\phi^{\rm eq} - \left(\frac{Y_\phi}{Y_\phi^{\rm eq}}\right)^{l-1} Y_\phi\right] = 0 \, , 
\end{equation}
where we have introduced the dimensionless variable $x \equiv m/T$, with $m$ denoting a generic mass scale, $g_{*s}(T)$ is the effective number of degrees of freedom in entropy\footnote{In our analysis we adopt the parametrization of $g_*(T)$ and $g_{*s}(T)$ derived in \cite{Saikawa:2018rcs}.} and $Y_\phi^{\rm eq}(x) = 45 \zeta(3) g_{n\phi} / (2\pi^4g_{*s}(x))$, with
\begin{equation}
	g_{n\phi} = g_\phi \times \begin{cases}
		1 \qquad {\rm bosons} \, , \\
		3/4 \quad {\rm fermions} \, .
		\end{cases}
\end{equation}
Here, $g_\phi$ denotes the number of internal degrees of freedom of the species $\phi$.

The light species $\phi$ contributes to the total energy density of radiation in the Universe. This is usually parameterized as
\begin{equation}
    \rho_\mathrm{rad} = \rho_\gamma \left[ 1 + \frac{7}{8}\left(\frac{4}{11}\right)^{4/3} N_{\rm eff} \right] \, ,
\end{equation}
where $\rho_\gamma = (\pi^2/15) T^4$ is the energy density of CMB photons and $N_{\rm eff}$ is the effective number of relativistic species, which can be decomposed as
\begin{equation}
	N_\mathrm{eff} \equiv N_{\rm eff}^{\rm SM} + \Delta N_\mathrm{eff} \, .
\end{equation}
The relic density of $\phi$ determines its contribution to $\Delta N_{\rm eff}$, which can be computed as \cite{DEramo:2021lgb} 
\begin{equation}
	\label{eq:DNeff}
    \Delta N_{\rm eff} = \frac{4}{7} \left(\frac{11}{4}\right)^{4/3} g_{*\phi} \left[ \frac{\frac{2\pi^4}{45\zeta(3)}g_{*s}^{\rm SM}(T_{\rm CMB})\frac{Y_\phi(T_{\rm CMB})}{g_{n\phi}}}{1-\frac{2\pi^4}{45\zeta(3)}g_{*\phi}\frac{Y_\phi(T_{\rm CMB})}{g_{n\phi}}} \right]^{4/3} \, ,
\end{equation}	
where $g_{*s}^{\rm SM}(T_{\rm CMB})= 2 + (7/11)N_{\rm eff}^{\rm SM} \simeq 3.94$ and 
\begin{equation}
	g_{*\phi} = g_\phi \times \begin{cases}
		1 \qquad {\rm bosons} \, , \\
		7/8 \quad {\rm fermions} \, .
		\end{cases}					 
\end{equation}

For each model that we consider in our analysis, we numerically integrate Eq. \eqref{eq:Boltzmann}, starting from different values of the reheating temperature, $T_{\rm reh}$.
As our initial condition, we assume a vanishing initial number density, i.e. $Y_\phi(x=x_{\rm reh}) = 0$. 
Then, we derive the contribution of the species under consideration to $\Delta N_{\rm eff}$ via Eq. \eqref{eq:DNeff} for the different reheating temperatures.

Before entering into the details of this analysis, we briefly discuss the expected contribution to $\Delta N_{\rm eff}$ both in the freeze-out and freeze-in regimes for a generic scenario. The transition between these two cases needs to be evaluated by numerically integrating the full Boltzmann equation \eqref{eq:Boltzmann}. This will be the subject of the rest of the paper, where we will focus on specific particle physics models.

\subsection{Thermal freeze-out regime}
Let's first briefly review what happens in the thermal regime, i.e. when the couplings of the species $\phi$ with SM fields are strong enough to bring $\phi$ in thermal equilibrium with the plasma. Then, at the temperature $T_d$ defined by the condition $\Gamma_\phi/H|_{T = T_d} = 1$, the light species decouples from the plasma.
The resulting contribution to $\Delta N_{\rm eff}$ is given by \cite{Brust:2013ova,Chacko:2015noa,Baumann:2016wac}
\begin{equation}
	\label{eq:DNeff-freeze-out}
	\Delta N_{\rm eff}^{\rm fo} \simeq 0.027 \times g_{*\phi} \left( \frac{g_{*s}^{\rm SM}(T_d)}{106.75} \right)^{-4/3} \, .
\end{equation}
This only depends on the decoupling temperature (and hence on the couplings of $\phi$ with SM particles), while there is no dependence on UV physics, such as the reheating temperature. The lowest contribution to $\Delta N_{\rm eff}$ is obtained when $\phi$ decouples above the mass of the heaviest SM particle (i.e., the top quark). This minimum amount of extra radiation reads
\begin{equation}
    \label{eq:DNeff-dof}
	\Delta N_{\rm eff}^{\rm fo} \simeq 
		\begin{cases}
			0.027 \quad \text{Goldstone boson (spin-$0$)} \, , \\
			0.047 \quad \text{Weyl fermion (spin-$1/2$)} \, , \\
			0.054 \quad \text{massless gauge boson (spin-$1$)} \, .
		\end{cases}
\end{equation}

\subsection{UV freeze-in regime}
In the regime where the couplings of $\phi$ with SM fields are not strong enough to bring the species in thermal equilibrium with the cosmological plasma, the production takes place via freeze-in \cite{Hall:2009bx}.
Considering a generic non-renormalizable operator of mass dimension $d = 4 + n$ (with $n \geq 1$) coupling $\phi$ to the SM, the production rate scales as \cite{Hall:2009bx,Elahi:2014fsa}
\begin{equation}
    \Gamma_\phi \propto \frac{T^{2n+1}}{\Lambda^{2n}} \, ,
\end{equation}
where $\Lambda$ is the cut-off energy scale of the effective field theory (EFT). In the freeze-in regime the dominant source term in the Boltzmann equation \eqref{eq:Boltzmann} is the one proportional to $Y_\phi^{\rm eq}$. Indeed, the production of the light species is dominated by high temperatures close to $T_{\rm reh}$, where, given the assumption of a vanishing initial density of $\phi$, it follows that $Y_\phi \ll Y_\phi^{\rm eq}$. Neglecting the temperature dependence of $g_*$ and $g_{*s}$, which are constant at high temperatures, we can straightforwardly integrate the Boltzmann equation to obtain the parametric dependence of $Y_\phi$ at the CMB epoch. Given that $H\propto T^2/M_{\rm Pl}$, this reads \cite{Elahi:2014fsa}
\begin{equation}
    Y_\phi(T_{\rm CMB}) \propto \int_{T_{\rm CMB}}^{T_{\rm reh}} dT \, \frac{M_{\rm Pl} T^{2n-2}}{\Lambda^{2n}} \, Y_\phi^{\rm eq} \propto  g_{n\phi} \frac{M_{\rm Pl} T_{\rm reh}^{2n-1}}{\Lambda^{2n}} \, ,
\end{equation}
where we have used the fact that $T_{\rm CMB} \ll T_{\rm reh}$.
From Eq. \eqref{eq:DNeff}, the contribution to $\Delta N_{\rm eff}$ in the UV freeze-in regime scales as 
\begin{equation}
    \Delta N_{\rm eff}^{\rm UV} \propto g_{*\phi} \left( \frac{M_{\rm Pl} T_{\rm reh}^{2n-1}}{\Lambda^{2n}} \right)^{4/3} \, .
\end{equation}
Note that $\Delta N_{\rm eff}^{\rm UV}$ depends not only on $\Lambda$ (which parametrizes the couplings of $\phi$ with SM fields), but also on the reheating temperature. Thus, an observational bound on the amount of extra radiation, $\Delta N_{\rm eff} < \delta$, translates into a lower limit on the EFT energy scale that also depends on $T_{\rm reh}$:
\begin{equation}
    \Lambda > \bar{\Lambda} \propto \left[ \left(\frac{g_{*\phi}}{\delta}\right)^{3/4} M_{\rm Pl} T_{\rm reh}^{2n-1} \right]^{{\frac{1}{2n}}} \, .
\end{equation}
Higher values of the reheating temperature imply stronger bounds on $\Lambda$, due to the increase in $\Delta N_{\rm eff}^{\rm UV}$.
Moreover, the higher the dimension of the non-renormalizable operator (larger $n$), the stronger the dependence on $T_{\rm reh}$.

\section{ALPs from Primakoff effect}
\label{sec:axions}
We begin by considering the case of ALPs, which are coupled to photons via the dimension-five operator
\begin{equation}
    \mathcal{L}_{a\gamma} = \frac{1}{4} g_{a\gamma} a F_{\mu \nu} \tilde{F}^{\mu\nu} \, ,
\end{equation}
where $F_{\mu\nu}$ is the photon field strength tensor and $\tilde{F}^{\mu\nu}=\epsilon^{\mu\nu\rho\sigma} F_{\rho\sigma}/2$ its dual.
ALPs can be produced in the primordial plasma from the resonant conversion of photons in the presence of the magnetic field of charged particles.
This is known as the Primakoff effect. The production rate for this process is \cite{Bolz:2000fu, Cadamuro:2011fd}
\begin{equation}
	\label{eq:primakoff}
	\Gamma_{\rm Prim} (T) \simeq \frac{\alpha_{\rm em} g^2_{a\gamma}}{36} g_q(T) \left[ \ln\left(\frac{T^2}{m_\gamma^2}\right) + 0.8194 \right] T^3 \, ,
\end{equation}
where $\alpha_{\rm em} = e^2/4\pi \simeq 1/137$ is the fine-structure constant, $m_\gamma = eT\sqrt{g_q(T)}/6$ is the photon plasmon mass and $g_q(T) = \sum_i q_i^2 g_{*,i}(T)$ is the effective number of relativistic charged degrees of freedom in the cosmological plasma, where $q_i$ denotes the charge of the $i$-th particle species. The evolution of $g_q$ as a function of $T$ is obtained as discussed in \cite{Caloni:2022uya}.
\begin{figure}
    \centering
    \includegraphics[width=0.7\textwidth]{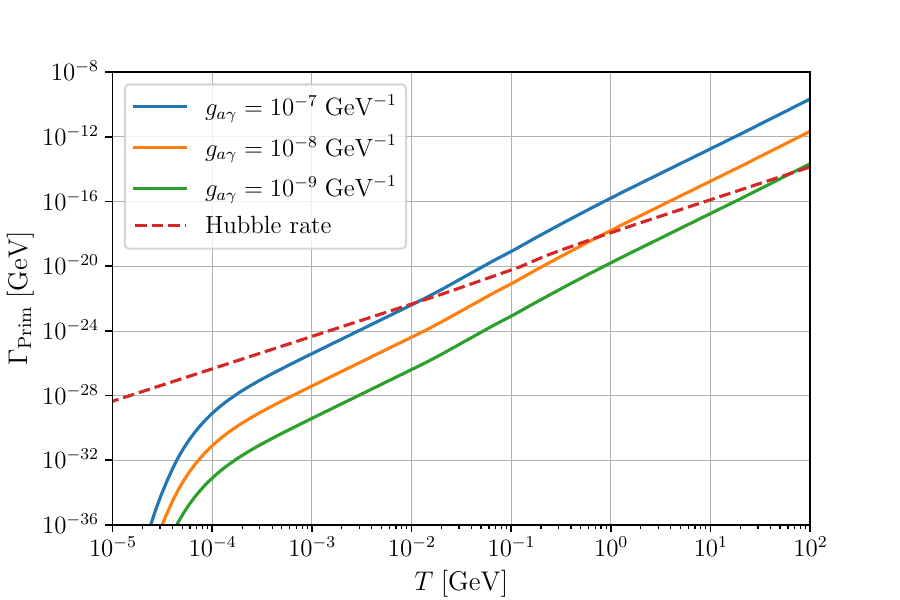}
    \caption{Production rate of ALPs from Primakoff effect as a function of the temperature $T$ (see Eq. \eqref{eq:primakoff}). The red dashed line represents the Hubble expansion rate.}
    \label{fig:ALPs-rate}
\end{figure}
The production rate from Primakoff effect as a function of the temperature is shown in Fig. \ref{fig:ALPs-rate}. This scales as $T^3$ at high temperatures and then gets Boltzmann suppressed when the temperature drops below the mass of the lightest SM charged particle, i.e. the electron. 

\begin{figure}
	\centering
	\includegraphics[width=0.7\textwidth]{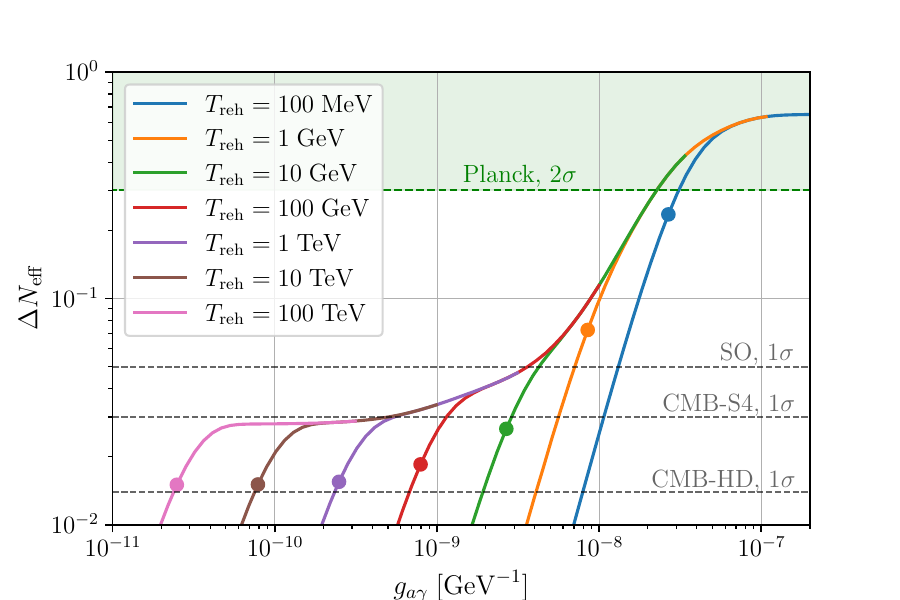}
	\caption{Contribution to $\Delta N_{\rm eff}$ from ALPs produced via Primakoff effect. Here, $\Delta N_{\rm eff}$ is shown as a function of the ALP-photon coupling, $g_{a\gamma}$, for different reheating temperatures. The green region is excluded by the bound derived from the Planck satellite, $\Delta N_{\rm eff} < 0.30$ at 95\% CL \cite{Planck:2018vyg}. We also show the forecasted sensitivity of CMB-S4 \cite{CMB-S4:2016ple}, SO \cite{SimonsObservatory:2018koc}, and of the futuristic CMB-HD \cite{CMB-HD:2022bsz}. The filled circles represent the values of $g_{a\gamma}$ for which $\Gamma_{\rm Prim}/H = 1$ at $T = T_{\rm reh}$, see the main text for more details.}
	\label{fig:axions-DNeff}
\end{figure}

Solving the Boltzmann equation \eqref{eq:Boltzmann} (with $l = 1$) for the production rate \eqref{eq:primakoff} and computing $\Delta N_{\rm eff}$ via Eq. \eqref{eq:DNeff}, we find the results shown in Fig. \ref{fig:axions-DNeff}. Here, we show the contribution of ALPs to $\Delta N_{\rm eff}$ as a function of the ALP-photon coupling $g_{a\gamma}$ for different values of the reheating temperature, $T_{\rm reh}$.
The green region is excluded by the bound on $\Delta N_{\rm eff}$ from the Planck satellite, $\Delta N_{\rm eff} < 0.30$ at 95\% CL \cite{Planck:2018vyg}.
We also show the forecasted sensitivity of CMB-S4 \cite{CMB-S4:2016ple} and of the futuristic CMB-HD \cite{CMB-HD:2022bsz}. The filled circles represent the values of $g_{a\gamma}$ for which $\Gamma_{\rm Prim}/H = 1$ at $T=T_{\rm reh}$. The part of each curve on the right of the circle represents the region of parameter space where ALPs are thermally produced via freeze-out. On the left of the circles we find instead the region of parameter space where ALPs never achieve thermal equilibrium with the cosmological plasma and are produced via freeze-in. Note that all the curves overlap in the thermal regime. 
Indeed, when the production takes place via freeze-out, the contribution to $\Delta N_{\rm eff}$ depends only on the temperature at which ALPs decouple from the primordial plasma. This is fixed by the strength of the ALP-photon coupling and does not depend on the value of the reheating temperature. For ALPs decoupling at temperatures above the top quark mass, we recover the well-known result reported in Eq. \eqref{eq:DNeff-dof}.

\begin{figure}
    \centering
    \includegraphics[width=0.7\textwidth]{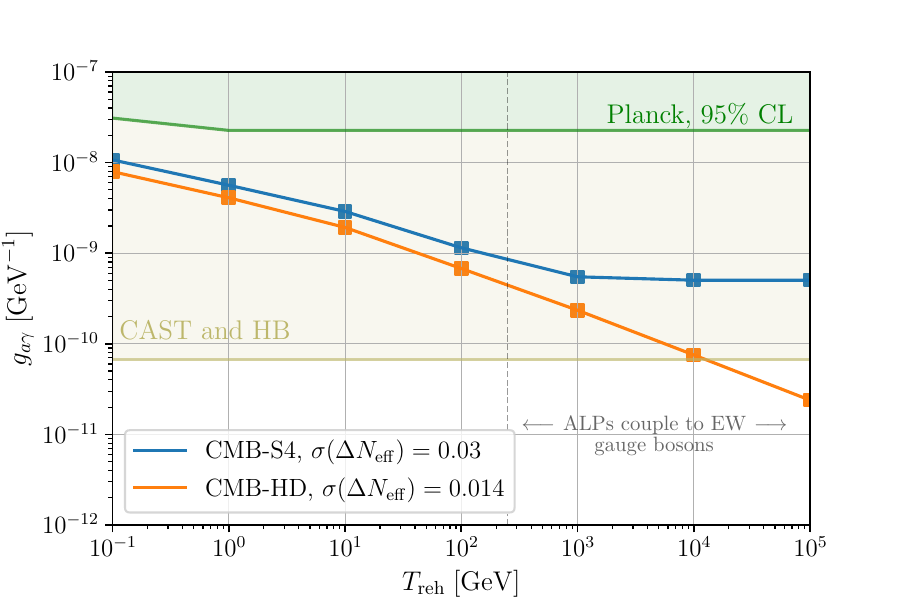}
    \caption{Forecasted $1\sigma$ upper bound on the ALP-photon coupling, $g_{a\gamma}$, as a function of the reheating temperature for the ground-based CMB-S4 experiment \cite{CMB-S4:2016ple} and for the futuristic CMB-HD \cite{CMB-HD:2022bsz}. The yellow shaded area is excluded by the limits derived from CAST \cite{CAST:2017uph} and from horizontal branch (HB) stars \cite{Ayala:2014pea}. Above the EW scale, $T_{\rm EW} \simeq 250 \; {\rm GeV}$, the ALP-photon coupling can be mapped to the coupling of ALPs with EW gauge bosons via Eq. \eqref{eq:ALP-aboveEW}. Note that for $T_{\rm reh} \ge 10 \; {\rm TeV}$ the bound on $g_{a\gamma}$ from CMB-S4 does not depend on the reheating temperature, since the threshold $\Delta N_{\rm eff} = 0.03$ is reached in the thermal regime (see Fig. \ref{fig:axions-DNeff}).}
    \label{fig:S4-axions}
\end{figure}

Note that, given the Planck sensitivity to $\Delta N_{\rm eff}$, for $T_{\rm reh} \gtrsim 1 \; {\rm GeV}$ we are able to probe only a region of parameter space where ALPs are produced via freeze-out, so that the limit on the ALP-photon coupling does not depend on the reheating temperature. On the other hand, for $T_{\rm reh} \simeq 100 \; {\rm MeV}$ the constraint on $\Delta N_{\rm eff}$ corresponds to a value of $g_{a \gamma}$ which is not large enough to bring ALPs in full thermal equilibrium with the cosmological plasma, since $\Gamma_{\rm Prim}/H \gtrsim 1$ at the initial temperature. This explains why the upper limit on $g_{a\gamma}$ is slightly relaxed. More precisely, we obtain the following limits on the ALP-photon coupling at 95\% CL:
\begin{equation}
    \begin{cases}
        g_{a\gamma} < 3.10 \times 10^{-8} \; {\rm GeV}^{-1}   \quad T_{\rm reh} = 100 \; {\rm MeV} \, , \\
        g_{a\gamma} < 2.27 \times 10^{-8} \; {\rm GeV}^{-1}  \quad T_{\rm reh} \ge 1 \; {\rm GeV} \, .
    \end{cases}
\end{equation}
Next-generation CMB experiments, like CMB-S4, will open a window into regions of parameter space where ALPs are produced via freeze-in. In this case, the limit on the ALP-photon coupling will be strongly dependent on the reheating temperature. This is shown in Fig. \ref{fig:S4-axions} for CMB-S4. 
Note that these values of the ALP-photon coupling are already excluded from the limits derived by the CAST helioscope at CERN \cite{CAST:2017uph} and from horizontal branch stars \cite{Friedland:2012hj,Ayala:2014pea}, which lead to an upper limit $g_{a\gamma} \lesssim 6.6 \times 10^{-11} \; {\rm GeV}^{-1}$. The futuristic CMB-HD would have enough sensitivity to $\Delta N_{\rm eff}$ to improve these limits for reheating temperatures $T_{\rm reh} \gtrsim 10 \; {\rm TeV}$. 

Note also that we are considering reheating temperatures up to 100 TeV. However, above the electroweak (EW) scale $T_{\rm EW} \simeq 250 \; {\rm GeV}$, when the EW symmetry is not broken yet, ALPs couple to $U(1)_Y$ and $SU(2)_L$ gauge bosons as (see e.g. \cite{Baumann:2016wac})
\begin{equation}
    \label{eq:L-above-EW}
    \mathcal{L}_{a\rm{EW}} = -\frac{1}{4}\frac{a}{\Lambda_a} \left( c_a B_{\mu\nu} \tilde{B}^{\mu\nu} + s_a W^a_{\mu\nu} \tilde{W}^{\mu\nu,a} \right) \, .
\end{equation}
Thus, for reheating temperatures $T_{\rm reh} > T_{\rm EW}$, the bounds on $g_{a\gamma}$ can be recast into bounds on the couplings to EW gauge bosons through the mapping \cite{Baumann:2016wac}
\begin{equation}
    \label{eq:ALP-aboveEW}
    g_{a\gamma} \rightarrow \frac{\cos^2\theta_w c_a + \sin^2\theta_w s_a}{\Lambda_a} \, ,
\end{equation}
where $\sin^2\theta_w \simeq 0.23$ \cite{CMS:2018ktx} is the sine of the Weinberg angle.

\section{Massless dark photons}
\label{sec:dark-photons}
Among the possible extensions of the SM of particle physics, one of the simplest possibility is to admit the existence of a new unbroken $U(1)$ gauge symmetry. The massless gauge boson of this new symmetry, which is usually referred to as dark photon (DP) (see Ref. \cite{Fabbrichesi:2020wbt} for a recent review),
does not couple to any of the SM currents, so that SM fields are neutral with respect to the new $U(1)$ symmetry. Moreover, no kinetic mixing with the ordinary photon is present for a massless DP \cite{Holdom:1985ag}. Nevertheless, the DP can still interact with SM particles via non-renormalizable operators \cite{Dobrescu:2004wz}. These interactions can play an important role in the early Universe, leading to a relic population of massless DPs that can significantly contribute to $\Delta N_{\rm eff}$ \cite{Salvio:2022hfa}.

We consider the effective Lagrangian up to dimension-six operators describing a massless DP coupled to SM fields \cite{Salvio:2022hfa, Dobrescu:2004wz}:
\begin{align}
	\nonumber
	\mathcal{L}_{\rm DP} &= - \frac{1}{4}P_{\mu\nu}P^{\mu\nu} + \frac{1}{M^2}P_{\mu\nu} \left(\bar{Q}_L\sigma^{\mu\nu}C_u\tilde{H}u_R + \bar{Q}_L\sigma^{\mu\nu}C_d H d_R + \bar{L}_L\sigma^{\mu\nu}C_e H e_R + {\rm h.c.}  \right) \\&
	+\frac{1}{M^2}P_{\mu\nu}H^{\dagger} \left(c_b B^{\mu\nu} + \tilde{c}_b \tilde{B}^{\mu\nu} + c_w W^{\mu\nu} + \tilde{c}_w \tilde{W}^{\mu\nu} + c_p P^{\mu\nu} + \tilde{c}_p \tilde{P}^{\mu\nu}\right) H \, ,
\end{align}
where $P_{\mu\nu} = \partial_\mu P_\nu - \partial_\nu P_\nu$ is the field strength tensor of the DP field, $P_\mu$, and $M$ is the EFT cut-off energy scale. The SM particle content includes the $SU(2)$ doublets
\begin{equation}
    Q_L = \begin{pmatrix}
			u_L \\
			d_L
		\end{pmatrix} \, , \quad
    L_L = \begin{pmatrix}
			\nu_L \\
			e_L
		\end{pmatrix} \, , 
\end{equation}
the singlets $u_R$, $d_R$, $e_R$, the Higgs doublet
\begin{equation}
    H = \begin{pmatrix}
			H^+ \\
			H^0
		\end{pmatrix} \, , \quad
  \tilde{H} \equiv i\sigma_2 H^* = \begin{pmatrix}
	                                   H^0 \\
	                                   H^+
	                                \end{pmatrix} \, ,
\end{equation}
and the EW gauge bosons $B_\mu$ and $W_\mu$, whose field strength tensors are $B_{\mu\nu}$ and $W_{\mu\nu}$, respectively.
Finally, $c_b$, $\tilde{c}_b$, $c_w$, $\tilde{c}_w$, $c_p$ and $\tilde{c}_p$ are real parameters, while $C_u$, $C_d$ and $C_e$ are complex $3\times 3$ matrices in flavor space. 

After the EW symmetry gets spontaneously broken, the Higgs field acquires a non-vanishing vacuum expectation value (VEV), $v_h = 246 \; {\rm GeV}$. Expanding the Higgs field around this VEV as
\begin{equation}
    \label{eq:Higgs-expansion}
	H = \frac{1}{\sqrt{2}}
		\begin{pmatrix}
			0 \\
			v_h+h(x
			)
		\end{pmatrix} \, , 
	\quad
		\tilde{H} = \frac{1}{\sqrt{2}}
	\begin{pmatrix}
	v_h+h(x) \\
	0
	\end{pmatrix} \, ,	
\end{equation}
we get the full set of operators in the broken EW phase.

In the upcoming sections we analyze the production of dark photons and compute the resulting contribution to $\Delta N_{\rm eff}$, considering the two cases in which DPs are produced either above or below the EW scale.

\subsection{Production above the electroweak phase transition}
Above the EW scale, where all the SM particles are effectively massless, the total production rate of DPs is obtained summing over the collection of scattering and annihilation processes described in Ref. \cite{Salvio:2022hfa}. 
\begin{figure}
    \centering
    \includegraphics[width=0.7\textwidth]{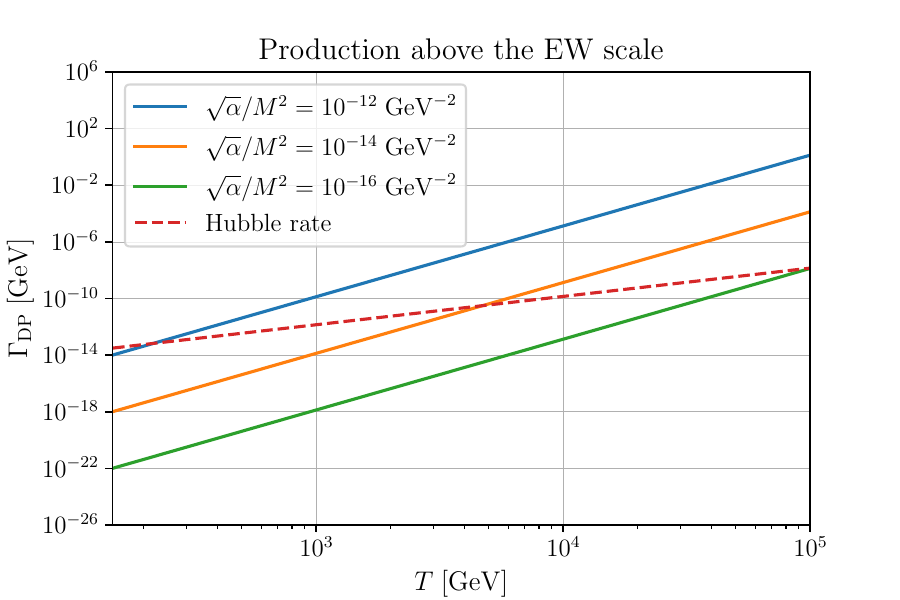}
    \caption{Production rate of dark photons above the EW scale. The red dashed line represents the Hubble expansion rate.}
    \label{fig:DP-rate-aboveEW}
\end{figure}
The total production rate is \cite{Salvio:2022hfa}
\begin{equation}
	\label{eq:Gamma_DP_aboveEW}
	\Gamma_{\rm DP} (T) = \frac{1}{2\pi \zeta(3)} \frac{\alpha}{M^4} T^5 \, ,
\end{equation}
where
\begin{equation}
    \label{eq:DP-alpha}
	\alpha \equiv \left[ 2.75C^2 + 1.31(c^2_b+\tilde{c}^2_b+c^2_w+\tilde{c}^2_w) + \frac{2\pi^6(c^2_p+\tilde{c}^2_p)}{675} \right] \, ,
\end{equation}
with
\begin{equation}
	C^2 \equiv 2 \left(3{\rm Tr}(C^\dagger_u C_u + C^\dagger_d C_d) + {\rm Tr}(C^\dagger_e C_e)\right) \, .
\end{equation}
This is shown as a function of the temperature in Fig. \ref{fig:DP-rate-aboveEW}.
Note that, given the steep dependence of the rate on the temperature ($\Gamma_{\rm DP} \propto T^5$), the production of DPs always takes place at temperatures very close to the reheating temperature. However, for large enough values of the effective coupling $\sqrt{\alpha}/M^2$, DPs thermally produced can still decouple from the primordial plasma at temperatures below $T_{\rm EW}$. Here, we focus on the case in which DPs either decouple at temperatures above $T_{\rm EW}$ or never achieve thermal equilibrium with the plasma. This is realized for couplings $\sqrt{\alpha}/M^2 \lesssim 10^{-12} \; {\rm GeV}^{-2}$, as can be seen in Fig. \ref{fig:DP-rate-aboveEW}.

\begin{figure}
    \centering
    \includegraphics[width=0.51\textwidth]{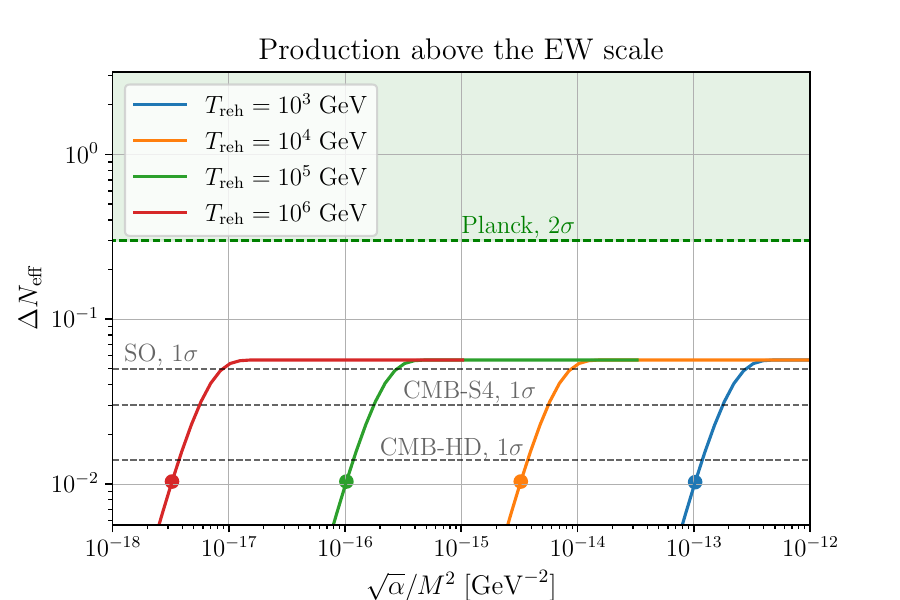} \hspace{-6mm}
    \includegraphics[width=0.51\textwidth]{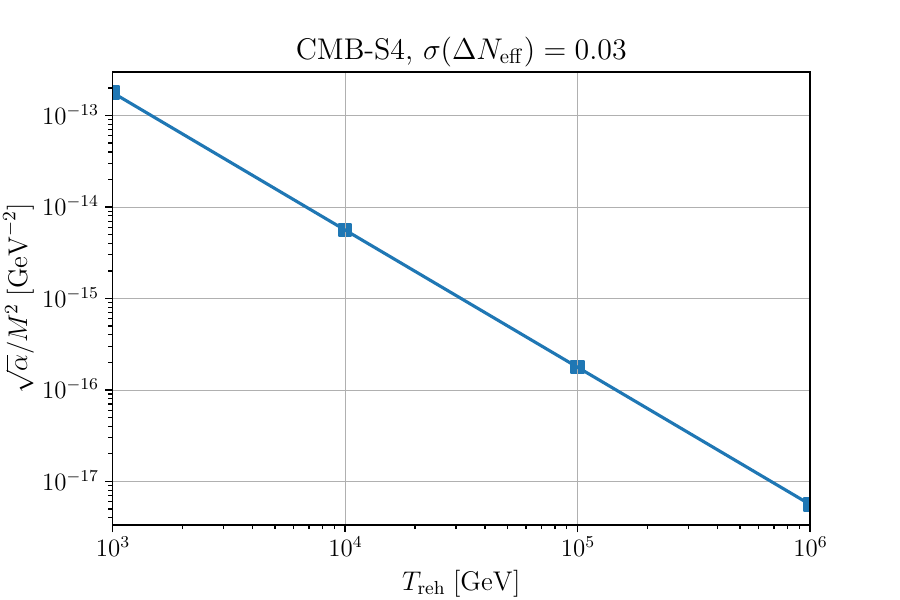}
    \caption{{\it Left panel}: $\Delta N_{\rm eff}$ as a function of the dark photon (DP) effective coupling, $\sqrt{\alpha}/M^2$, for different reheating temperatures $T_\mathrm{reh}$. Here, we are in the regime where the (would be) decoupling temperature is above the EW scale, corresponding to $\sqrt{\alpha}/M^2 \lesssim 10^{-12} \; {\rm GeV}^{-2}$. {\it Right panel}: forecasted $1\sigma$ upper bound on $\sqrt{\alpha}/M^2$ as a function of the reheating temperature for CMB-S4, assuming $\sigma(\Delta N_{\rm eff})_{\rm{CMB-}S4} = 0.03$ \cite{CMB-S4:2016ple}.}
    \label{fig:S4-DPs-aboveEW}
\end{figure}

We integrate the Boltzmann equation \eqref{eq:Boltzmann} with the rate in Eq. \eqref{eq:Gamma_DP_aboveEW} and then compute $\Delta N_{\rm eff}$ via Eq. \eqref{eq:DNeff}.\footnote{The production rate for DPs above the EW scale has been obtained in Ref. \cite{Salvio:2022hfa} by summing over different production channels, with either one ($C$, $c_b$, $\tilde{c}_b$, $c_w$ and $\tilde{c}_w$ couplings) or two ($c_p$ and $\tilde{c}_p$ couplings) DP particles in the final state. Thus, in principle we should separate the contributions with $l=1$ and $l=2$ in the Boltzmann equation \eqref{eq:Boltzmann}. However, as also discussed in Sec. \ref{sec:UV-freezein}, the dominant contribution for freeze-in production in the Boltzmann equation is the one proportional to the equilibrium density. Thus, we choose to keep a unique effective coupling $\sqrt{\alpha}/M^2$, defined via Eq. \eqref{eq:DP-alpha}, and solve the Boltzmann equation with $l = 1$.} This is shown as a function of $\sqrt{\alpha}/M^2$ is shown in Fig. \ref{fig:S4-DPs-aboveEW} (left panel) for different reheating temperatures above $T_{\rm EW}$. Note that the amount of extra radiation produced in the thermal regime is within the reach of SO at 1$\sigma$, but slightly below the $2\sigma$ sensitivity of CMB-S4. This is expected for a massless gauge boson which decouples from the primordial plasma at temperatures above the top quark mass, see Eq. \eqref{eq:DNeff-dof}.
The $1\sigma$ sensitivity of CMB-S4 on $\Delta N_{\rm eff}$ will instead allow us to probe the regime in which DPs are produced via freeze-in. The corresponding bound on $\sqrt{\alpha}/M^2$ as a function of the reheating temperature is shown in the right panel of Fig. \ref{fig:S4-DPs-aboveEW}. 

\subsection{Production below the electroweak phase transition}
\label{sec:dark-photons-belowEW}
After the electroweak phase transition, the Higgs field has to expanded around its VEV as in Eq. \eqref{eq:Higgs-expansion}. The Lagrangian describing the interactions between the DP and SM fermions leads to both dimension-5 and dimension-6 operators:
\begin{equation}
	\mathcal{L}_{\rm DP, ferm}^{<{\rm EW}} = \mathcal{L}^{\rm dim-5}_{\rm DP, ferm} + \mathcal{L}^{\rm dim-6}_{\rm DP, ferm} \, .
\end{equation}
In the mass eigenstate basis we have
\begin{align}
	\mathcal{L}^{\rm dim-5}_{\rm DP, ferm} &= \frac{v_h}{\sqrt{2}M^2} \sum_{f=u,d,e} P_{\mu\nu} \Big[ \bar{f} \sigma^{\mu\nu}\left(d_m^f + i \gamma_5 d_e^f\right)f \Big] \, , \\
	\mathcal{L}^{\rm dim-6}_{\rm DP, ferm} &= \frac{1}{\sqrt{2}M^2} \sum_{f=u,d,e} P_{\mu\nu} \Big[ \bar{f} \sigma^{\mu\nu}\left(d_m^f + i \gamma_5 d_e^f\right)h f \Big] \, ,
\end{align}
where
\begin{align}
	d_e^f &= {\rm Re} \left(U_L^f C_f U_R^{f\dagger}\right) \equiv {\rm Re} C_f' \, , \\
	d_m^f &= {\rm Im} \left(U_L^f C_f U_R^{f\dagger}\right)  \equiv {\rm Im} C_f' 
\end{align}
are ``dark'' electric and magnetic dipole moments of SM fermions, $f = u, d, e$. In the above expressions, $U_L^f$ and $U_R^f$ are the unitary matrices that diagonalize the masses of the fermion $f$. 

In this work, we focus on the production of DPs from SM leptons ($\ell = e, \mu, \tau$) via the dimension-5 dipole operators. This has been first studied in Ref. \cite{Dobrescu:2004wz} for muons in the instantaneous decoupling approximation. We improve this analysis by integrating the Boltzmann equation for the number density of DPs and calculating the associated contribution to $\Delta N_{\rm eff}$. This also allows us to extend the analysis of \cite{Dobrescu:2004wz} to the freeze-in regime. Moreover, we also analyze the contributions from electrons and taus.
The relevant processes for DP production are lepton annihilations ($\ell\bar{\ell}\rightarrow\gamma A$) and Compton-like scatterings ($\ell\gamma\rightarrow \ell A$), where $A$ denotes the DP. The cross-section for these interactions is given by \cite{Dobrescu:2004wz} 
\begin{equation}
    \sigma^{\rm DP}_{\rm dipole} \simeq \alpha_{\rm em}\frac{v_h^2}{2} \frac{|(C'_e)_{ii}|^2}{M^4} \, .
\end{equation}
Here, $|(C'_e)_{ii}|$ are the universal flavor couplings between the DP and SM leptons, with $i = 1, 2, 3$ for electrons, muons and taus, respectively. We can factorize out the dependence on the mass of SM leptons (which enters via the respective Yukawa couplings) by rewriting $|(C'_e)_{ii}|/M^2$ as
\begin{equation}
    \label{eq:DPs-mass-hierachy}
    \frac{|(C'_e)_{ii}|}{M^2} = \frac{1}{\Lambda_M^2} y_i =  \frac{1}{\Lambda_M^2} \left( \frac{\sqrt{2}m_i}{v_h}\right) \, .
\end{equation}
This explicitly shows the mass hierarchy between the couplings of the DP with SM leptons,  $|(C'_e)_{33}| \gg |(C'_e)_{22}| \gg |(C'_e)_{11}|$. We show results for benchmarks with either flavor universal couplings or couplings which incorporate the mass hierarchy of the SM leptons.

\begin{figure}
    \centering
    \includegraphics[width=0.7\textwidth]{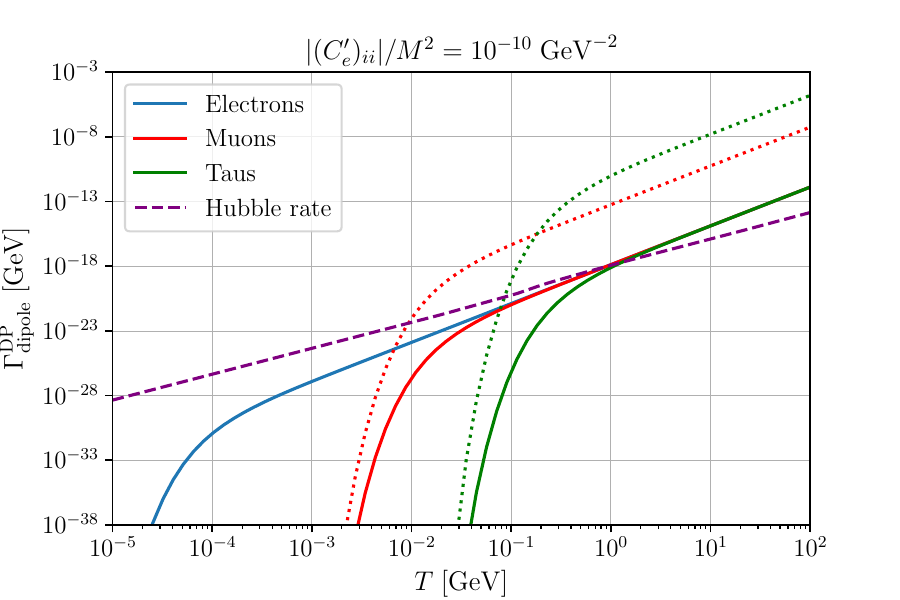}
    \caption{Production rate of dark photons via ``dark'' dipole operators below the EW scale. The solid lines represent the contributions from electrons (blue), muons (red) and taus (green) for fixed universal couplings, $|(C'_e)_{ii}|/M^2 = 10^{-10} \; {\rm GeV}^{-2}$. The dotted lines correspond to the contributions from muons (red) and taus (green) factorizing out the mass hierarchy dependence from the couplings (see Eq. \eqref{eq:DPs-mass-hierachy}) such that the electron coupling is normalized to $|(C'_e)_{11}|/M^2 = 10^{-10} \; {\rm GeV}^{-2}$.}
    \label{fig:DP-rate-belowEW}
\end{figure}

\begin{figure}
    \centering
    \includegraphics[width=0.51\textwidth]{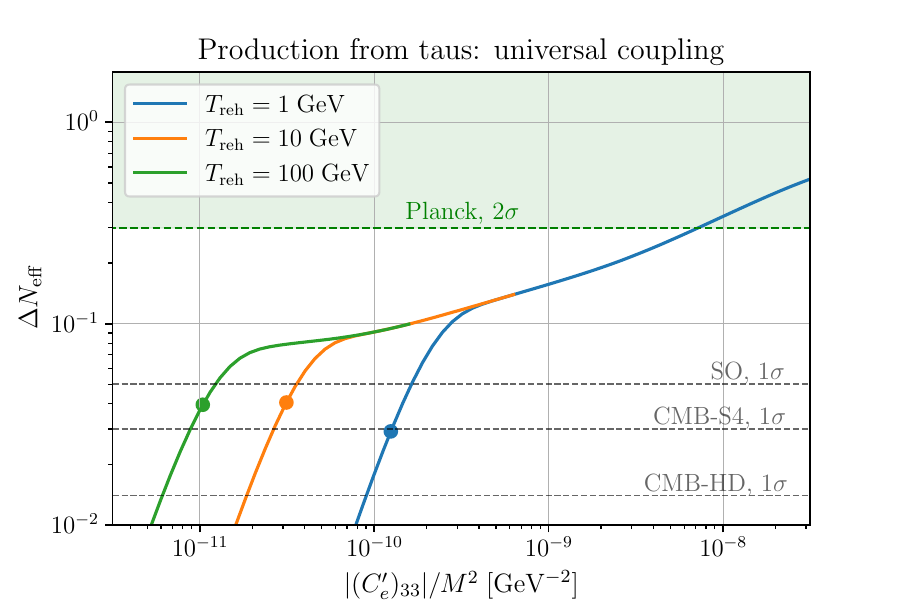} \hspace{-6mm}
    \includegraphics[width=0.51\textwidth]{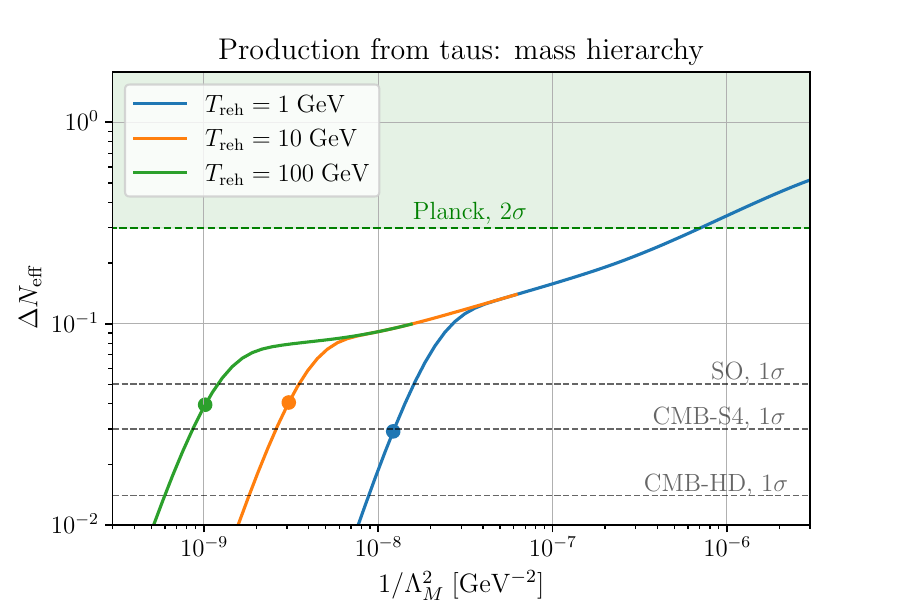}
    \includegraphics[width=0.51\textwidth]{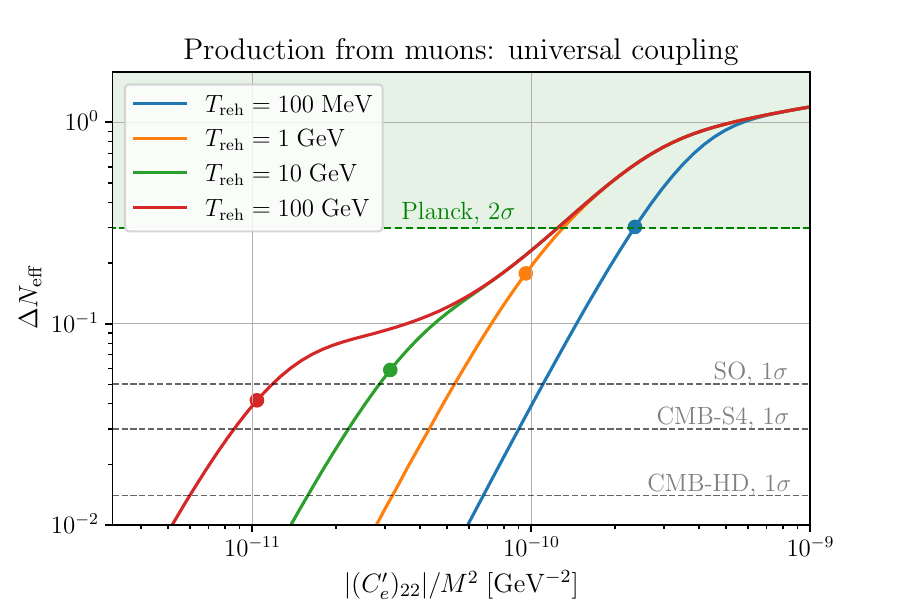} \hspace{-6mm}
    \includegraphics[width=0.51\textwidth]{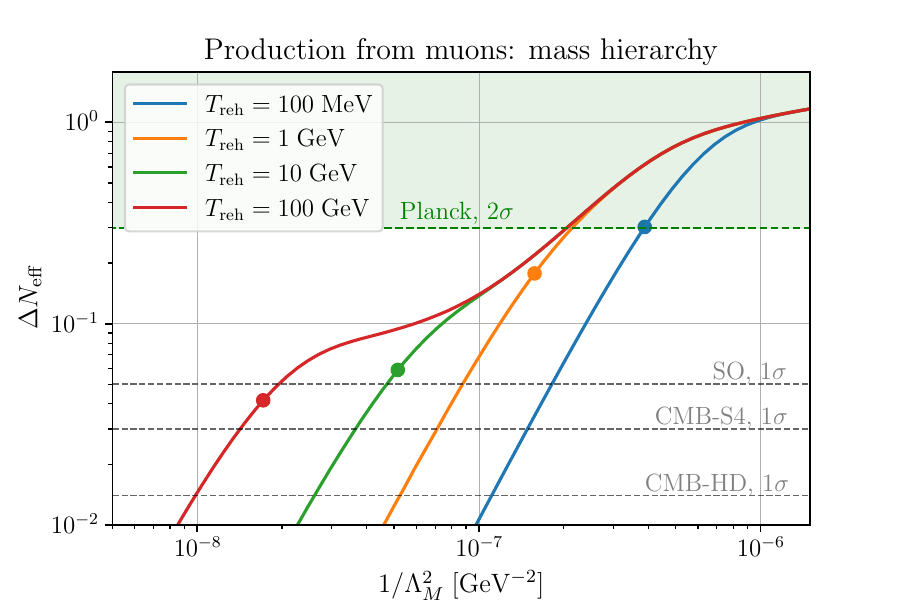} 
    \includegraphics[width=0.51\textwidth]{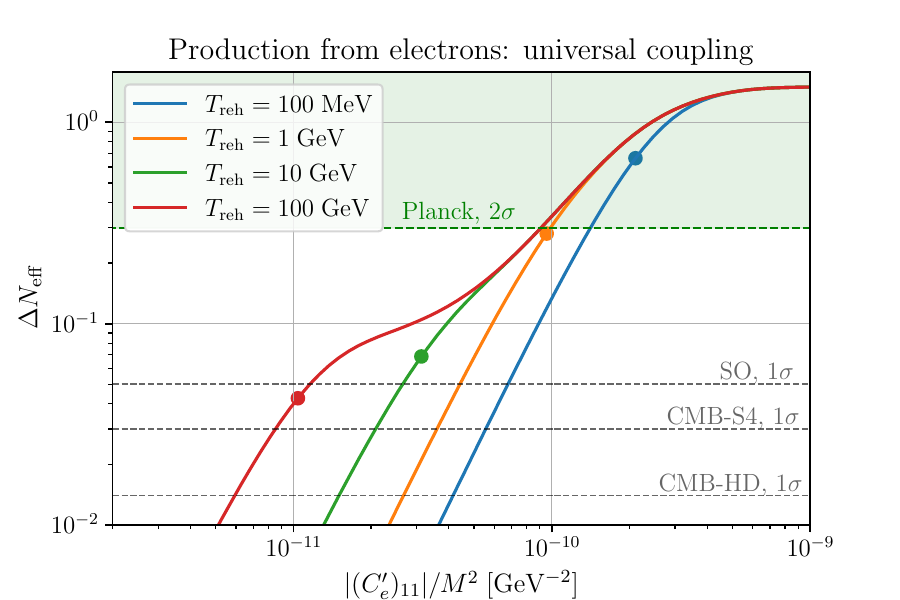} \hspace{-6mm}
    \includegraphics[width=0.51\textwidth]{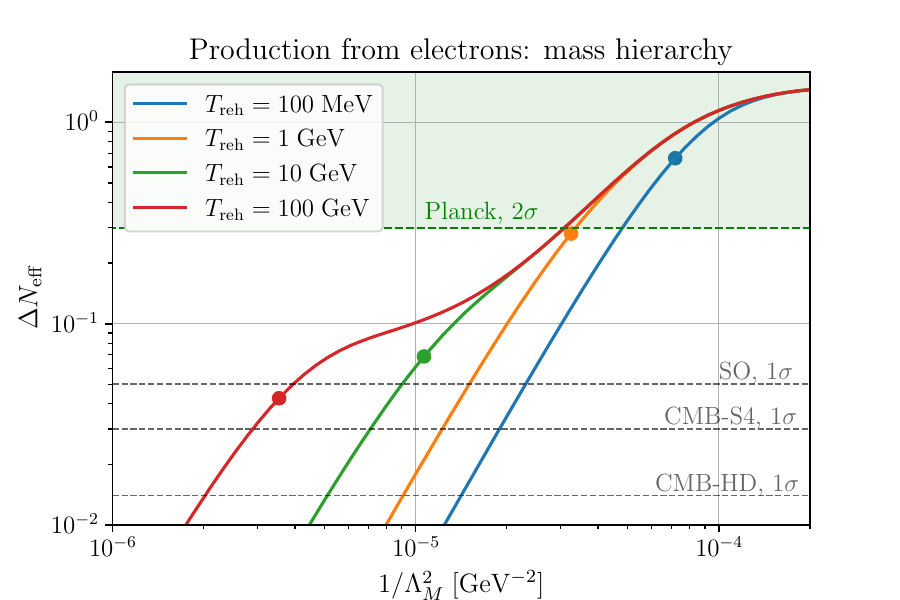} 
    \caption{$\Delta N_{\rm eff}$ from dark photons produced via dipole operators below the EW scale. In the left panels we show the contribution to $\Delta N_{\rm eff}$ from taus (upper panel), muons (central panel) and electrons (lower panel) as a function of the corresponding universal couplings, $|(C'_e)_{ii}|/M^2$. The right panels show the same contributions to $\Delta N_{\rm eff}$ in terms of $1/\Lambda_M^2$, thus factorizing out the mass hierarchy suppression in the couplings between dark photons and leptons.}
    \label{fig:DNeff-DP-belowEW}
\end{figure}

\begin{figure}
    \centering 
    \includegraphics[width=0.51\textwidth]{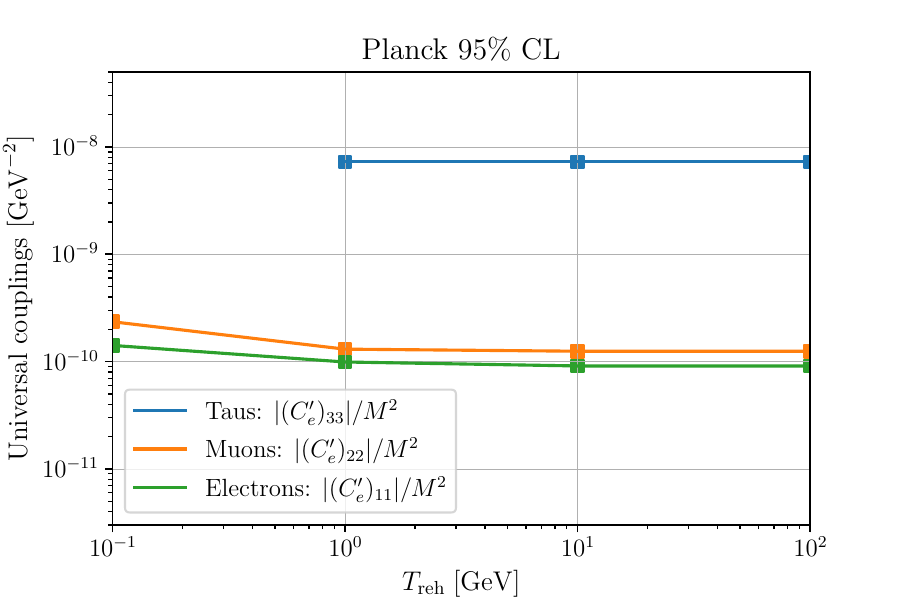} \hspace{-6mm}
    \includegraphics[width=0.51\textwidth]{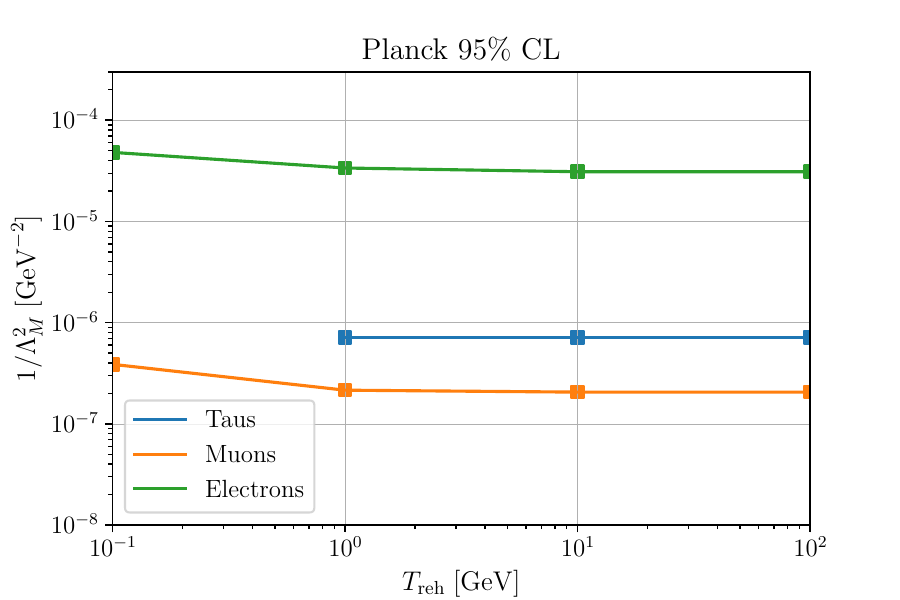} 
    \includegraphics[width=0.51\textwidth]{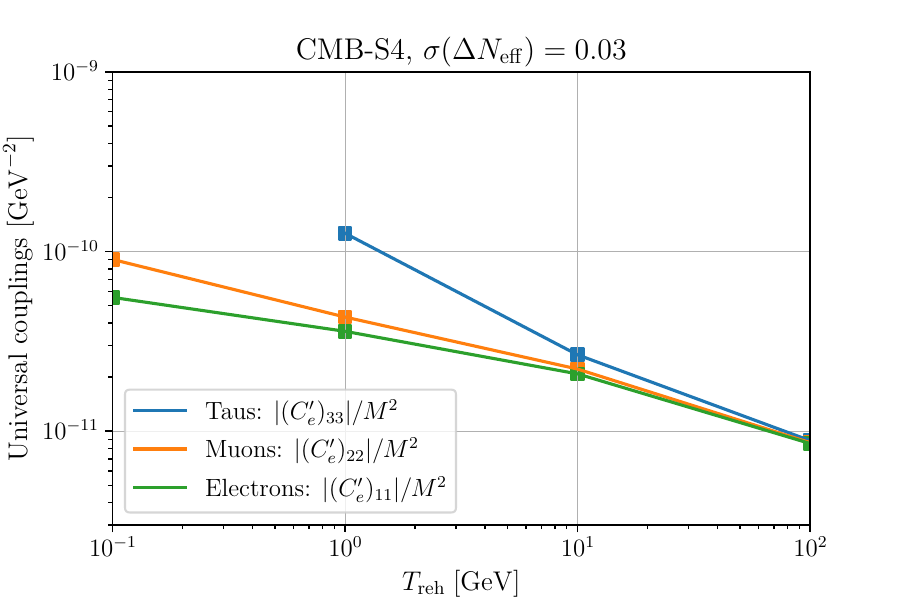} \hspace{-6mm}
    \includegraphics[width=0.51\textwidth]{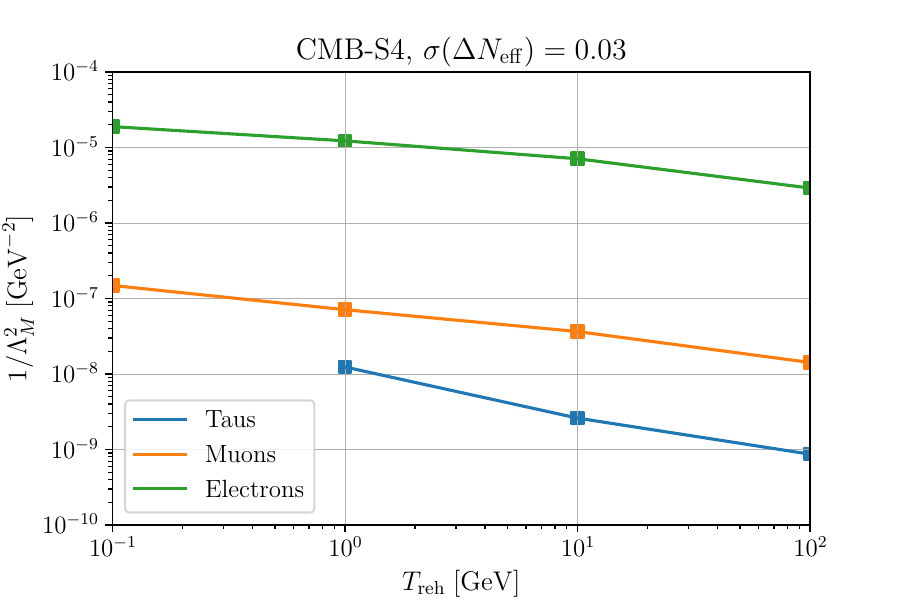} 
    \caption{Constraints on the couplings between the DP and SM leptons from Planck at 95\% CL in terms of either the universal couplings (upper left panel) or $1/\Lambda_M^2$ (upper right panel). The lower panels show the forecasted 1$\sigma$ limits for CMB-S4.}
    \label{fig:DP-belowEW-constraints}
\end{figure}

Turning to the production rates, we first consider lepton-antilepton annihilations. 
The thermally-averaged DP production rate can be computed as
\begin{equation}
    \label{eq:DPbelow-rate-ann}
    \Gamma_{\ell\bar{\ell}\rightarrow\gamma A}^{\rm DP} = \langle \sigma_{\ell\bar{\ell}\rightarrow\gamma A}^{\rm DP} v_{\rm rel} \rangle \frac{n_\ell^{\rm eq}n_{\bar{\ell}}^{\rm eq}}{n_A^{\rm eq}} = \frac{g_\ell g_{\bar{\ell}}}{n_A^{\rm eq}} \int \frac{d^3p}{(2\pi)^3} \frac{d^3k}{(2\pi)^3} f_\ell^{\rm eq}(E_\ell) f_{\bar{\ell}}^{\rm eq}(E_{\bar{\ell}}) \sigma^{\rm DP}_{\rm dipole} v_{\rm rel} \, ,
\end{equation}
where $\mathbf{p}$ and $\mathbf{k}$ are the three-momenta of the colliding leptons and antileptons, having energies $E_\ell = (|\mathbf{p}|^2+m_\ell^2)^{1/2}$ and $E_{\bar{\ell}} = (|\mathbf{k}|^2+m_\ell^2)^{1/2}$, respectively. The equilibrium distribution function for leptons is $f_\ell^{\rm eq}(E_\ell) = [\exp(E_\ell/T) + 1]^{-1}$ and 
\begin{equation}
	v_{\rm rel} = \frac{s}{2E_\ell E_{\bar{\ell}}} \sqrt{1-\frac{4m_\ell^2}{s}}
\end{equation}
is the relative (M\o{}ller) velocity between the colliding particles in the center of momentum frame. Here, 
\begin{equation}
	s = 2m_\ell^2 + 2 E_\ell E_{\bar{\ell}} - 2pk\cos\theta
\end{equation}
is the squared center of mass energy, where $\theta$ denotes the angle between the two colliding SM leptons.
In order to proceed with the computation, it is useful to introduce the following quantities \cite{Gondolo:1990dk}
\begin{equation}
    E_+ \equiv E_\ell + E_{\bar{\ell}} \, , \quad E_- \equiv E_\ell - E_{\bar{\ell}} \, ,
\end{equation}
so that the rate in Eq. \eqref{eq:DPbelow-rate-ann} can be written as
\begin{align}
    \label{eq:DPbelow-rate-ann-s}
    \nonumber
    \Gamma_{\ell\bar{\ell}\rightarrow\gamma A}^{\rm DP} &= \frac{|(C'_e)_{ii}|^2}{M^4} \frac{\alpha_{\rm em} v_h^2}{32 \pi^4 n_A^{\rm eq}}  \int_{4m_\ell^2}^{\infty} ds \int_{\sqrt{s}}^{\infty} dE_+ \int_{y_-}^{y_+} dE_- \left[\exp\left(\frac{E_+ + E_-}{2T}\right) + 1\right]^{-1} \\
    & \times \left[\exp\left(\frac{E_+ - E_-}{2T}\right) + 1\right]^{-1} s \sqrt{1-\frac{4m_\ell^2}{s}} \, ,
\end{align}
with
\begin{equation}
    y_{\mp} \equiv \mp \sqrt{1-\frac{4m_\ell^2}{s}} \sqrt{E_+^2-s} \, .
\end{equation}
Under the approximation of a Maxwell-Boltzmann distribution for leptons in the thermal bath, we can analytically integrate over $E_+$ and $E_-$, so that Eq. \eqref{eq:DPbelow-rate-ann-s} reduces to
\begin{equation}
    \Gamma_{\ell\bar{\ell}\rightarrow\gamma A}^{\rm DP}(T) = \frac{|(C'_e)_{ii}|^2}{M^4} \frac{\alpha_{\rm em} v_h^2}{16\pi^2 \zeta(3) T^2} \int_{4m_\ell^2}^\infty \sqrt{s}\left(s-4m_\ell^2\right) K_1\left(\frac{\sqrt{s}}{T}\right) ds \, , 
\end{equation}
where $K_1$ is the modified Bessel function of order 1 and we have used the fact that
\begin{equation}
    n_A^{\rm eq} = g_A \int \frac{d^3k}{(2\pi)^3} f_A^{\rm eq}(E_A) = \frac{2\zeta(3)}{\pi^2} T^3 \, .
\end{equation}
Analogously, the production rate of DPs from Compton-like scattering reads
\begin{equation}
    \Gamma_{\ell\gamma\rightarrow \ell A}^{\rm DP}(T) = \Gamma_{\bar{\ell}\gamma\rightarrow \bar{\ell} A}^{\rm DP}(T) = \frac{|(C'_e)_{ii}|^2}{M^4} \frac{\alpha_{\rm em} v_h^2}{32\pi^2 T^2} \int_{m_\ell^2}^\infty \sqrt{s}\left(s-m_\ell^2\right) K_1\left(\frac{\sqrt{s}}{T}\right) ds \, . 
\end{equation}
The total production rate of DPs via ``dark'' dipole operators is
\begin{equation}
    \Gamma_{\rm dipole}^{\rm DP} (T) = \Gamma_{\ell\bar{\ell}\rightarrow\gamma A}^{\rm DP} (T) + 2 \Gamma_{\ell\gamma\rightarrow \ell A}^{\rm DP}(T) \, .
\end{equation}
This is shown in Fig. \ref{fig:DP-rate-belowEW} for electron (blue), muon (red) and tau (green) processes. The rate scales as $T^3$ in the relativistic regime and then gets Boltzmann suppressed when the temperature drops below the lepton masses.
We can then integrate the Botzmann equation \eqref{eq:Boltzmann} for DPs (with $l = 1$, since both lepton annihilations and Compton-like scatterings yield one DP particle in the final state) and compute the contribution to $\Delta N_{\rm eff}$ via Eq. \eqref{eq:DNeff}. This is shown in Fig. \ref{fig:DNeff-DP-belowEW} as a function of either the universal flavor couplings (left panels) or factorizing out the mass dependence as in Eq. \eqref{eq:DPs-mass-hierachy} (right panels). 
Note that the freeze-in contributions reflect the mass hierarchy suppression of the couplings. However, in the thermal regime the contribution to $\Delta N_{\rm eff}$ from taus is more suppressed with respect to that of muons, contrary to what we might naively expect. This is due to the Boltzmann suppression of the production rate when the temperature drops below the tau mass (see Fig. \ref{fig:DP-rate-belowEW}). Indeed, the limit $\Delta N_{\rm eff} < 0.30$ from Planck translates into a bound on the decoupling temperature $T_d \gtrsim 100 \; {\rm MeV}$. Since the number density of taus is Boltzmann suppressed below $T \sim 1 \; {\rm GeV}$, we need a relatively large universal coupling between taus and DPs in order keep the latter in thermal equilibrium with the plasma until $T \sim 100 \; {\rm MeV}$. This also explains why the contribution to $\Delta N_{\rm eff}$ from DP-tau interactions are suppressed for a reheating temperature of 100 MeV. 
Fig. \ref{fig:DP-belowEW-constraints} summarizes the constraints on the DP-lepton couplings from Planck at 95\% CL. We also show the forecasted 1$\sigma$ limits for CMB-S4, assuming $\sigma(\Delta N_{\rm eff}) = 0.03$.

\section{Right-handed neutrinos in \texorpdfstring{$B-L$}{} model}
\label{sec:sterile-neutrinos-B-L}
\begin{figure}
	\centering
    \includegraphics[width=0.7\textwidth]{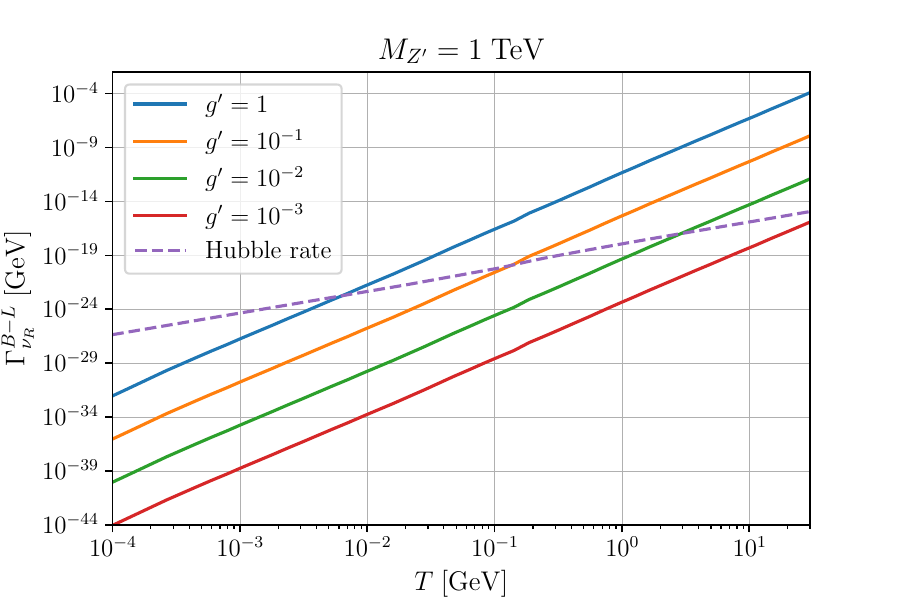}
	\caption{Total production rate of right-handed neutrinos in the $B-L$ model in the regime where $M_{Z'} \gg T$. In this plot we show the case with $M_{Z'} = 1 \; {\rm TeV}$. The purple dashed line represents the Hubble expansion rate.}
	\label{fig:RHN-rate-MB}
\end{figure}

In this section we consider an extension of the SM model which features a $U(1)_{B-L}$ gauge symmetry. This model includes three right-handed neutrinos and a new gauge boson $Z'$ of mass $M_{Z'}$. The Lagrangian describing the interactions of the $Z'$ boson with SM fermions and the right-handed neutrinos is \cite{Langacker:1991pg,Barger:2003zh,Heeck:2014zfa}
\begin{equation}
	\mathcal{L} = g' Z'_\mu \sum_i \left[ \frac{1}{3}\left(\bar{u}_i\gamma^\mu u_i + \bar{d}_i\gamma^\mu d_i\right) - \bar{e}_i\gamma^\mu e_i - \bar{\nu}_{L,i}\gamma^\mu \nu_{L,i} - \bar{\nu}_{R,i}\gamma^\mu \nu_{R,i}  \right] \, ,
\end{equation}
where the index $i$ runs over the generations of SM fermions. Right-handed neutrinos can be produced in the early Universe from the annihilations of SM fermions mediated by the $Z'$ boson: 
\begin{equation}
    \label{eq: RHN-process}
 	f + \bar{f} \rightarrow \nu_R + \bar{\nu}_R \, .
\end{equation}
The cross-section for this process is given by \cite{Adshead:2022ovo}
\begin{equation}
	\label{eq:xsection-B-L}
	\sigma_{f\bar{f}\rightarrow\nu_R\bar{\nu}_R}^{B-L} = \frac{Q^2_{B-L}(f) N_c(f)g'^4}{2\pi} \frac{s}{(s-M_{Z'}^2)^2+\Gamma_{Z'}^2M_{Z'}^2} \sqrt{\frac{s}{s-4m_f^2}} \left(1+\frac{2m_f^2}{s}\right) \, ,
\end{equation}
where $Q_{B-L}(f)$ is the $B-L$ charge of the SM fermion $f$ of mass $m_f$, $N_c(f)$ its color number and $\Gamma_{Z'}$ is the total decay width of the $Z'$ boson, which is given by \cite{Adshead:2022ovo}
\begin{equation}
	\Gamma_{Z'} = \frac{g'^2}{12\pi} M_{Z'} \left[ 3 + \sum_{f}^{2m_f<M_{Z'}}Q^2_{B-L}(f) N_c(f)\left(1+\frac{2m_f^2}{M_{Z'}^2}\right) \sqrt{1-\frac{4m_f^2}{M_{Z'}^2}} \right] \, .
\end{equation} 
To compute the production rate of right-handed neutrinos, it is convenient to rewrite the cross-section \eqref{eq:xsection-B-L} using the narrow width approximation
\begin{equation}
    \label{eq:NWA}
	\frac{s}{(s-M_{Z'}^2)^2+\Gamma_{Z'}^2M_{Z'}^2} \approx \frac{s}{M_{Z'}^4} \Theta(M_{Z'}^2-s) + \frac{\pi M_{Z'}}{\Gamma_{Z'}} \delta(s-M_{Z'}^2) + \frac{1}{s}\Theta(s-M_{Z'}^2) \, .
\end{equation}
Note that the resonant part of the cross-section, which corresponds to the term with the Dirac delta function, is proportional to $g'^2$ due to the presence of $\Gamma_{Z'}$ in the denominator, while the contributions arising from contact interactions are proportional to $g'^4$ (and hence are more suppressed, given that $g' \ll 1$).
In our analysis we focus on the regime with $M_{Z'} \gg T$, where the $Z'$ boson can be integrated out. This leads to an effective four-fermion interaction which, being a dimension-six operator, results in a production rate that scales as $T^5$ \cite{Barger:2003zh,Heeck:2014zfa}. Thus, we consider only the first term in Eq. \eqref{eq:NWA} in the computation of the production rate. This follows the steps outlined in section \ref{sec:dark-photons-belowEW}, so that we find 
\begin{align}
    \nonumber
    \label{eq:RHNs-rate-MB}
    \Gamma_{f\bar{f}\rightarrow\nu_R\bar{\nu}_R}^{B-L} (T) = \langle \sigma_{f\bar{f}\rightarrow\nu_R\bar{\nu}_R}^{B-L} v_{\rm rel} \rangle \frac{n^{\rm eq}_f n^{\rm eq}_{\bar{f}}}{n^{\rm eq}_{\nu_R}} &= \frac{g'^4}{M_{Z'}^4} \frac{Q_{B-L}^2(f) N_c(f)}{24\pi^3\zeta(3)T^2} \int_{4m_f^2}^{\infty} ds \sqrt{s-4m^2_f} \left(s^2+2m_f^2 s\right) \\
    & \times K_1\left(\frac{\sqrt{s}}{T}\right) \Theta(M_{Z'}^2-s)  \, .
\end{align}
The Heaviside function implies that the integral in Eq. \eqref{eq:RHNs-rate-MB} is non-vanishing only if $M_{Z'}^2 > 4m_f^2$. In this case, the upper limit of integration is $M_{Z'}^2$.
The total rate is obtained summing Eq. \eqref{eq:RHNs-rate-MB} over all SM fermions:
\begin{equation}
    \Gamma_{\nu_R}^{B-L}(T) = \sum_f \Gamma_{f\bar{f}\rightarrow\nu_R\bar{\nu}_R}^{B-L} (T) \, .
\end{equation}
We include free quarks ($Q_{B-L} = 1/3$) in the sum for $T > 150 \; {\rm MeV}$, while for $T < 150 \; {\rm MeV}$ we sum only over SM leptons ($f = e, \mu, \tau, \nu_L$), for which $Q_{B-L} = -1$. The massess of SM fermions are taken from \cite{Workman:2022ynf}.
In Fig. \ref{fig:RHN-rate-MB} we show the total production rate of RH$\nu$s for $M_{Z'} = 1 \; {\rm TeV}$. As anticipated (see also Ref. \cite{Heeck:2014zfa}), the rate scales as $T^5$ in the regime where $M_{Z'} \gg T$.

\begin{figure}
    \centering
    \includegraphics[width=0.51\textwidth]{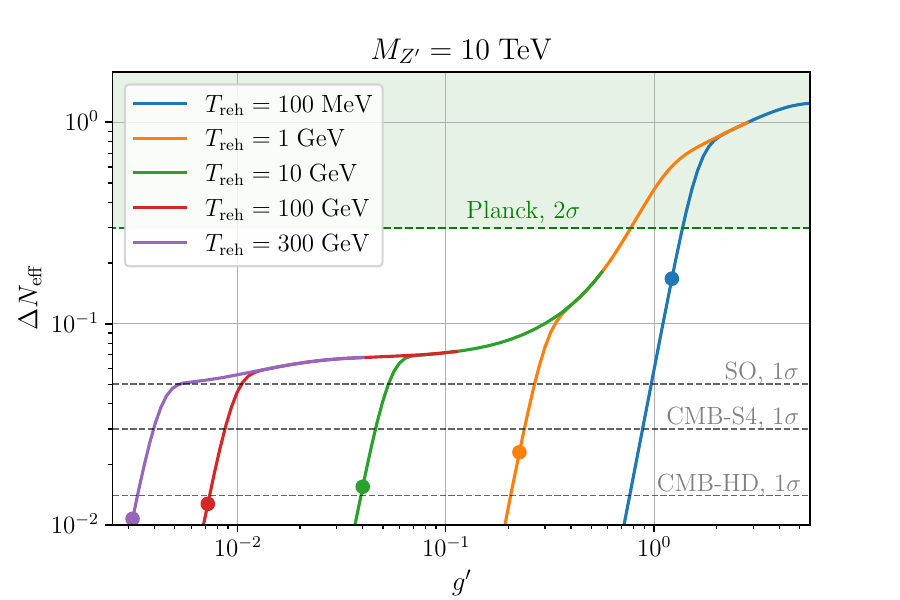} \hspace{-6mm}
    \includegraphics[width=0.51\textwidth]{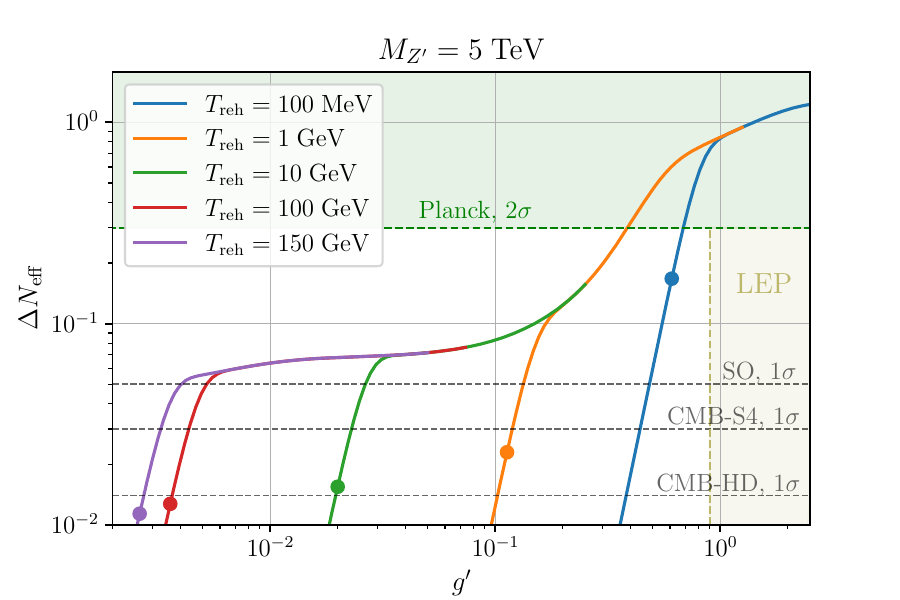}
     \includegraphics[width=0.51\textwidth]{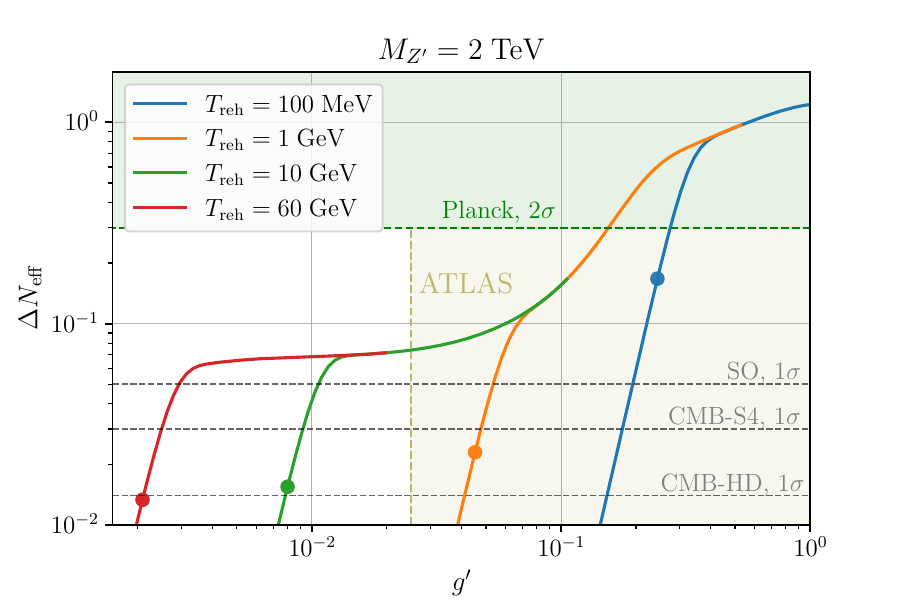} \hspace{-6mm}
    \includegraphics[width=0.51\textwidth]{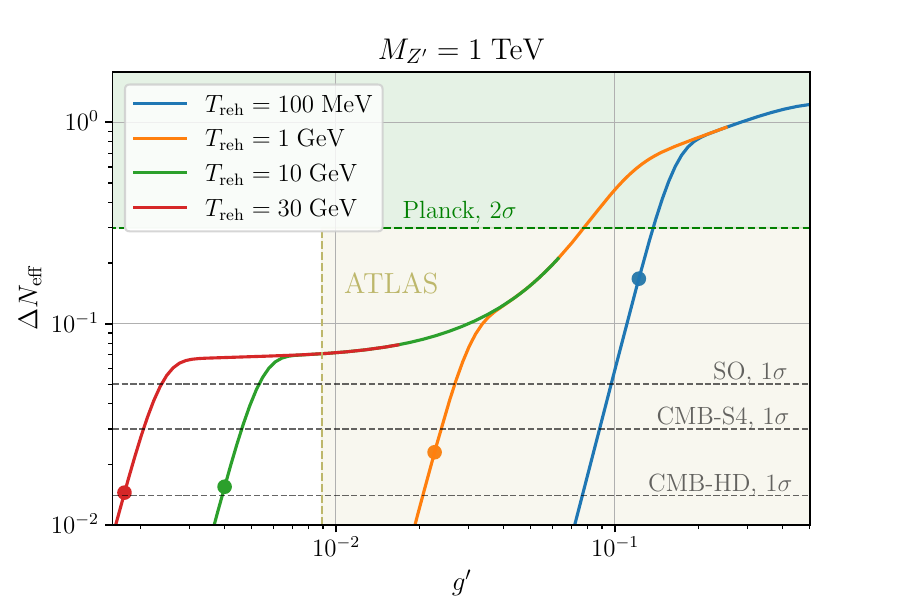} 
    \includegraphics[width=0.51\textwidth]{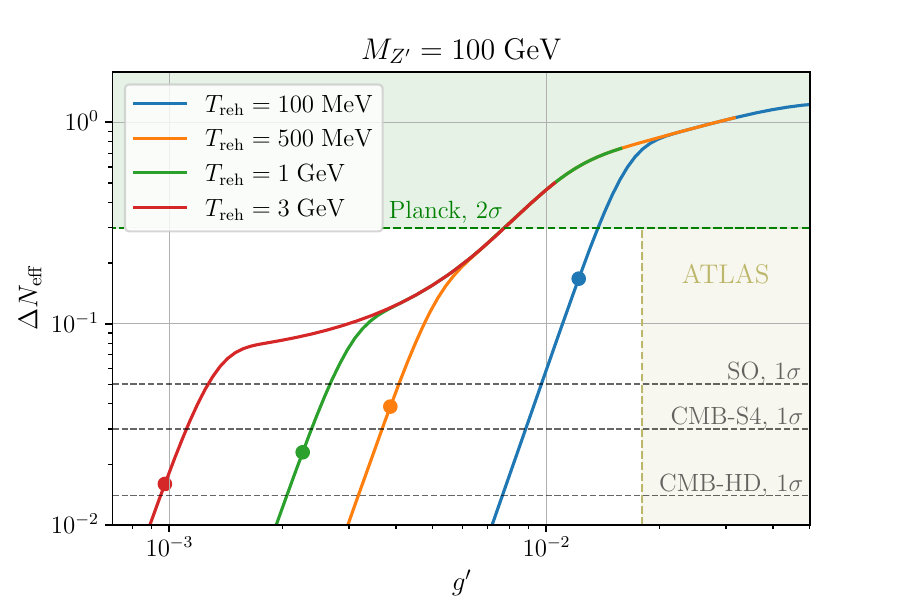}
    \caption{$\Delta N_{\rm eff}$ from right-handed neutrinos in the $B-L$ model. This is shown as a function of the coupling $g'$ for different reheating temperatures and masses of the $Z'$ boson, in the regime where $M_{Z'} \gg T$. The filled circles represent the values of $g'$ for which $\Gamma/H = 1$ at $T = T_{\rm reh}$ (see the main text for more details). 
    The green region is excluded by the bound derived from the Planck satellite \cite{Planck:2018vyg}, while the yellow region in each plot is excluded by the limits derived by ATLAS \cite{ATLAS:2017fih,ATLAS:2017rue,Escudero:2018fwn}. In particular, the bound for the case with $M_{Z'} = 100 \; {\rm GeV}$ is derived by analyzing the
    modification to the Drell-Yan differential cross section (see \cite{Escudero:2018fwn}) using measurements from ATLAS at $\sqrt{s} = 8 \; {\rm TeV}$ with a luminosity of $20.2 \; {\rm fb}^{-1}$ \cite{ATLAS:2017rue}. Note that there are no constraint on $g'$ from collider experiments for $M_{Z'} = 10 \; {\rm TeV}$. We also show the forecasted sensitivity of SO \cite{SimonsObservatory:2018koc}, CMB-S4 \cite{CMB-S4:2016ple} and of the futuristic CMB-HD \cite{CMB-HD:2022bsz}.}
    \label{fig:DNeff-RHN-B-L}
\end{figure}

\begin{figure}
    \centering
    \includegraphics[width=0.515\textwidth]{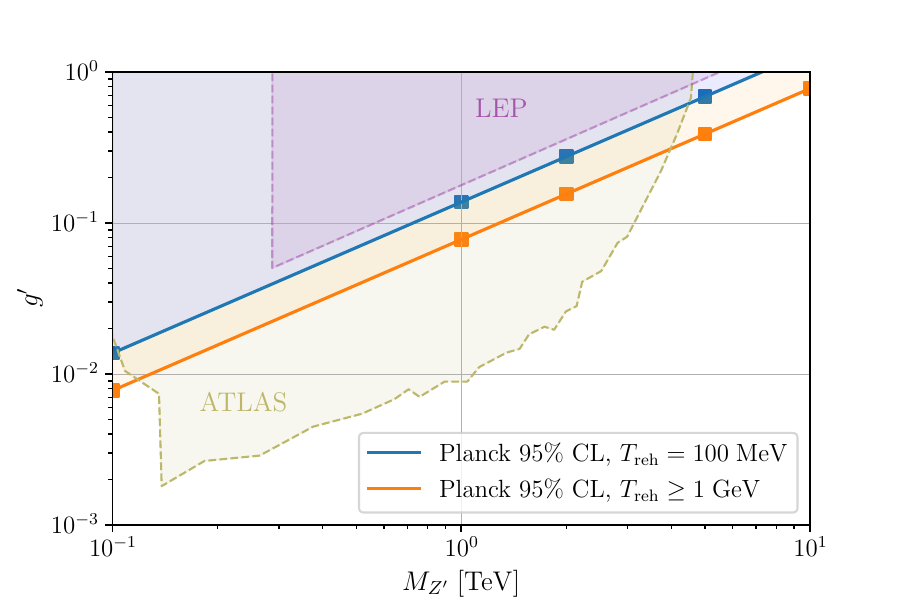} \hspace{-8mm}
    \includegraphics[width=0.515\textwidth]{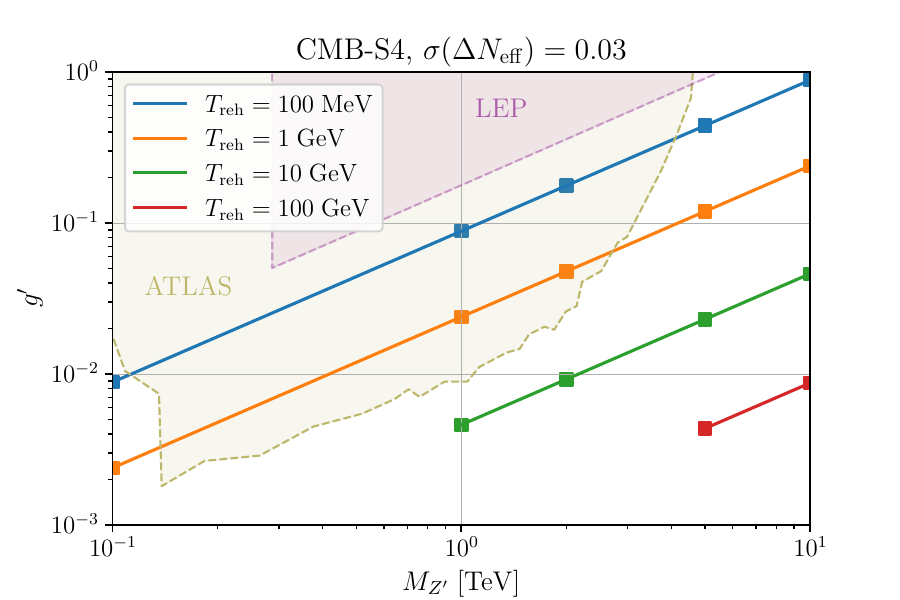}
    \caption{\textit{Left panel}: Constraints on $g'$ as a function of the mass of the $Z'$ boson, $M_{Z'}$, from Planck. The yellow region is excluded by the limits derived by ATLAS \cite{ATLAS:2017fih,ATLAS:2017rue,Escudero:2018fwn}, while the purple region is excluded by LEP \cite{Escudero:2018fwn,LEP:2004xhf}. The contours are adapted from Ref. \cite{Escudero:2018fwn}. \textit{Right panel}: Forecasted $1\sigma$ upper bound on $g'$ as a function of $M_{Z'}$ for CMB-S4. This is shown for different reheating temperatures from $100 \; {\rm MeV}$ to $100 \; {\rm GeV}$.}
    \label{fig:RHN-S4}
\end{figure}

We can now solve the Boltzmann equation \eqref{eq:Boltzmann} (with $l = 2$) and compute $\Delta N_{\rm eff}$ via Eq. \eqref{eq:DNeff}. This is shown in Fig. \ref{fig:DNeff-RHN-B-L} for masses of the $Z'$ boson $M_{Z'} = 1, 2, 5, 10 \; {\rm TeV}$ and $M_{Z'} = 100 \; {\rm GeV}$. The constraints derived on $g'$ from Planck at 95\% CL are also shown in the left panel of Fig. \ref{fig:RHN-S4}.
We compare our limits with those derived from collider experiments. For masses of the $Z'$ boson $M_{Z'} = 1, 2 \; {\rm TeV}$, the constraints obtained from the ATLAS collaboration \cite{Escudero:2018fwn,ATLAS:2017fih} are stronger than the cosmological ones, while for $M_{Z'} = 5 \; {\rm TeV}$ the cosmological limit on $g'$ is stronger than the one from colliders, which comes from LEP \cite{Escudero:2018fwn,LEP:2004xhf,Appelquist:2002mw,Carena:2004xs}.
Also for $M_{Z'} = 100 \; {\rm GeV}$, the cosmological constraints are stronger than the one from colliders, that are derived by analyzing the modification to the Drell-Yan differential cross section using measurements from ATLAS at $\sqrt{s} = 8 \; {\rm TeV}$ with a luminosity of $20.2 \; {\rm fb}^{-1}$ \cite{Escudero:2018fwn,ATLAS:2017rue}. Finally, for $M_{Z'} = 10 \; {\rm TeV}$ there are no constraints from collider experiments, meaning that at the moment cosmology represents a unique probe to explore this region of parameter space. 

In the right panel of Fig. \ref{fig:RHN-S4}, we show how the constraints on $g'$ will be improved by CMB-S4, assuming $\sigma(\Delta N_{\rm eff}) = 0.03$ \cite{CMB-S4:2016ple}.
Note that for reheating temperatures $T_{\rm reh} \gtrsim 10 \; {\rm GeV}$, the limits on $g'$ from CMB-S4 will be stronger than those obtained from collider experiments even for masses of the $Z'$ boson of 1 and 2 TeV.

Before closing this section, we want to point out that, even if we have focused on regions of parameter space satisfying the condition $M_{Z'} \gg T$, we can assume that our limits are valid even for reheating temperatures of order or higher than the $Z'$ mass. In this case, our bounds on $g'$ have to be interpreted as conservative limits. Indeed, any process taking place at temperatures $\gtrsim M_{Z'}$ should result in a relic density of RH$\nu$s that adds up to $\Delta N_{\rm eff}$, thus implying stronger bounds on $g'$. This consideration agrees with the results derived in Ref. \cite{Adshead:2022ovo}.

\section{Charge radius of right-handed neutrinos}
\label{sec:chargeradius}
The study of neutrino properties is one of the most active research fields in high-energy physics, since it represents a powerful window into new physics beyond the SM. A particularly interesting task is the characterization (both experimental and theoretical) of the possible electromagnetic properties of neutrinos \cite{Giunti:2014ixa}. One of these is the neutrino charge radius, $\langle r_\nu^2 \rangle$, which is predicted to be non-zero within the SM \cite{Bernabeu:2004jr} and can receive additional contributions from BSM physics.

If neutrinos are Dirac particles, fermion-antifermion annihilations can produce right-handed neutrino-antineutrino pairs ($f\bar{f} \rightarrow \nu_R \bar\nu_R$) through the coupling induced by the neutrino charge radius of right-handed neutrinos, $\langle r_\nu^2 \rangle_R$. Considering only electron-positron annihilations, this process has been used to constrain $\langle r_\nu^2 \rangle_R$ from SN1987A energy loss considerations, resulting in the limit $\langle r_\nu^2 \rangle_R < 2 \times 10^{-33} \; {\rm cm}^2$ \cite{Grifols:1989vi}, and from the impact of $\Delta N_{\rm eff}$ on Big Bang Nucleosynthesis, which gives $\langle r_\nu^2 \rangle_R < 7 \times 10^{-33} \; {\rm cm}^2$ under the approximation of instantaneous decoupling \cite{Grifols:1986ed}. 

Laboratory constraints on the neutrino charge radius of SM neutrinos are summarized in Ref. \cite{Giunti:2014ixa}. Note however that left-handed and right-handed neutrinos have in general a different charge radius and the relation between these two quantities is model dependent (see e.g. \cite{Grifols:1989vi}). This means that the limits on $\langle r_\nu^2 \rangle_R$ can not be used to infer bounds on $\langle r_\nu^2 \rangle_L$ in a model independent way. 

\begin{figure}
	\centering
	\includegraphics[width=0.7\textwidth]{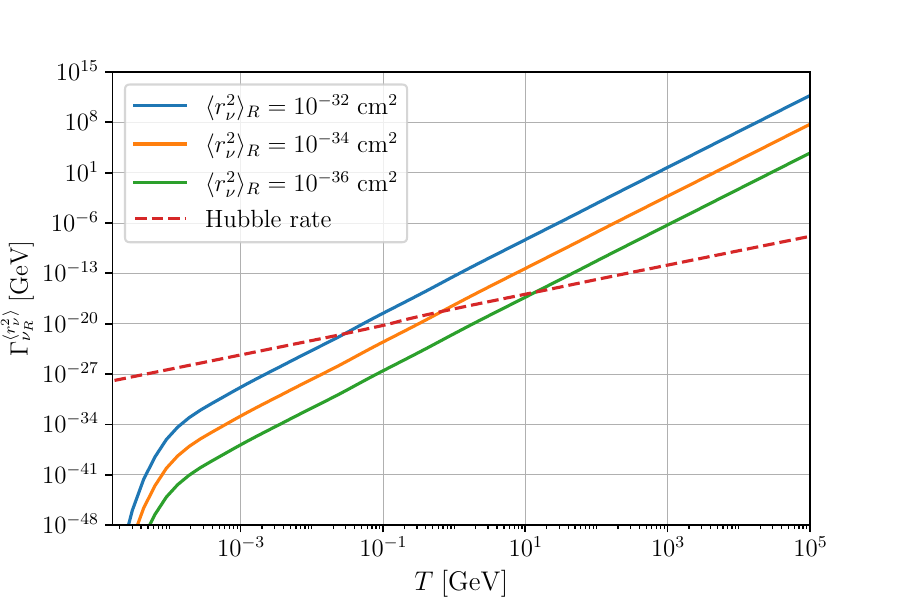}
	\caption{Total production rate of right-handed neutrinos as a function of the temperature $T$ for different values of the charge radius $\langle r_\nu^2 \rangle_R$. The red dashed line represents the Hubble expansion rate.}
	\label{fig:Rate-charge-radius}
\end{figure}

In this section we re-examine the contribution of light right-handed neutrinos to $\Delta N_{\rm eff}$, going beyond the simple thermal regime and accounting for the contributions from all SM fermions.
The cross-section for the process under consideration is \cite{Grifols:1986ed, Grifols:1989vi}
\begin{equation}
    \sigma_{f\bar{f} \rightarrow \nu_R \bar\nu_R}^{\langle r_{\nu}^2 \rangle} = \frac{\pi q_f^4 \alpha_{\rm em}^2 N_c(f)}{54} \langle r_{\nu}^2 \rangle^2_R s \, ,
\end{equation}
where $q_f$ is the electric charge of the fermion $f$ in units of the elementary charge.
Following the steps already outlined in Sec. \ref{sec:sterile-neutrinos-B-L}, we can then compute the production rate of right-handed neutrinos. Under the approximation of a Maxwell-Boltzmann distribution for SM fermions in the thermal bath, we find 
\begin{equation}
    \label{eq:RHN-rate-MB}
    \Gamma_{f\bar{f}\rightarrow\nu_R\bar{\nu}_R}^{\langle r_{\nu}^2 \rangle} (T) = \frac{q_f^4\alpha_{\rm em}^2 N_c(f)}{162 \pi \zeta(3)}  \frac{\langle r_\nu^2\rangle^2_R}{T^2} \int_{4m_f^2}^{\infty} ds \, \sqrt{s} \left(s^2-4m_f^2 s\right) K_1\left(\frac{\sqrt{s}}{T}\right) \, .
\end{equation}
The total production rate is obtained summing Eq. \eqref{eq:RHN-rate-MB} over SM leptons and quarks:
\begin{equation}
    \label{eq:rate-charge-radius}
    \Gamma_{\nu_R}^{\langle r_{\nu}^2 \rangle} (T) = \sum_f \Gamma_{f\bar{f}\rightarrow\nu_R\bar{\nu}_R}^{\langle r_{\nu}^2 \rangle} (T) \, ,
\end{equation}
where the contributions of free quarks is included only for temperatures $T > 150 \; {\rm MeV}$.
We show $\Gamma_{\nu_R}^{\langle r_{\nu}^2 \rangle}$ as a function of the temperature in Fig. \ref{fig:Rate-charge-radius} for different values of $\langle r_{\nu}^2 \rangle_R$. Note that the rate scales as $T^5$ in the relativistic regime and then gets suppressed when the temperature drops below the mass of the lightest fermion, i.e. the electron.

\begin{figure}
	\centering
	\includegraphics[width=0.7\textwidth]{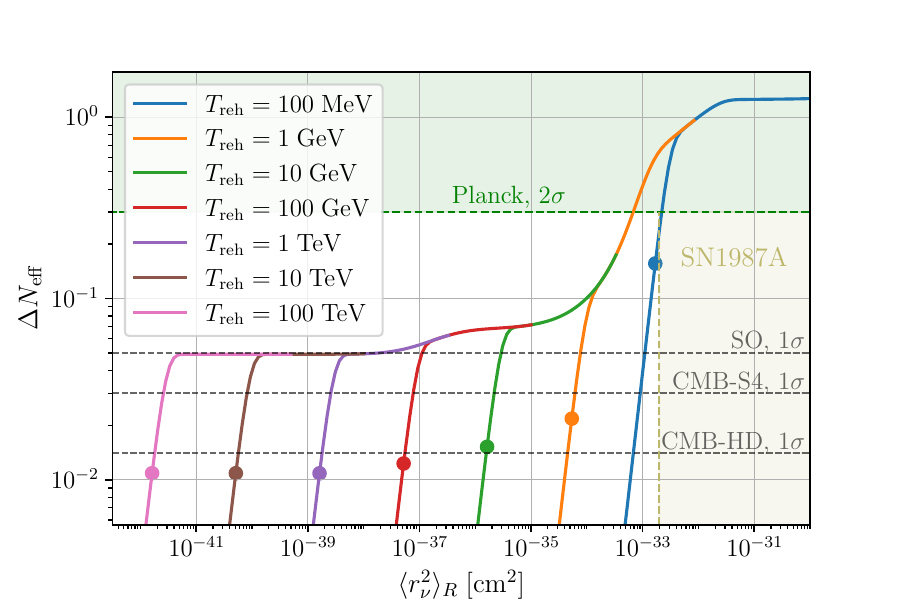}
	\caption{$\Delta N_{\rm eff}$ from right-handed neutrinos produced by fermion-antifermion annihilations induced by the neutrino charge radius, $\langle r_\nu^2 \rangle_R$. The filled circles represent the values of $\langle r_\nu^2 \rangle_R$ for which $\Gamma/H = 1$ at $T = T_{\rm reh}$. The yellow shaded region is excluded by SN1987A energy loss considerations \cite{Grifols:1989vi}.}
	\label{fig:DNeff-charge radius}
\end{figure}

\begin{figure}
    \centering
    \includegraphics[width=0.7\textwidth]{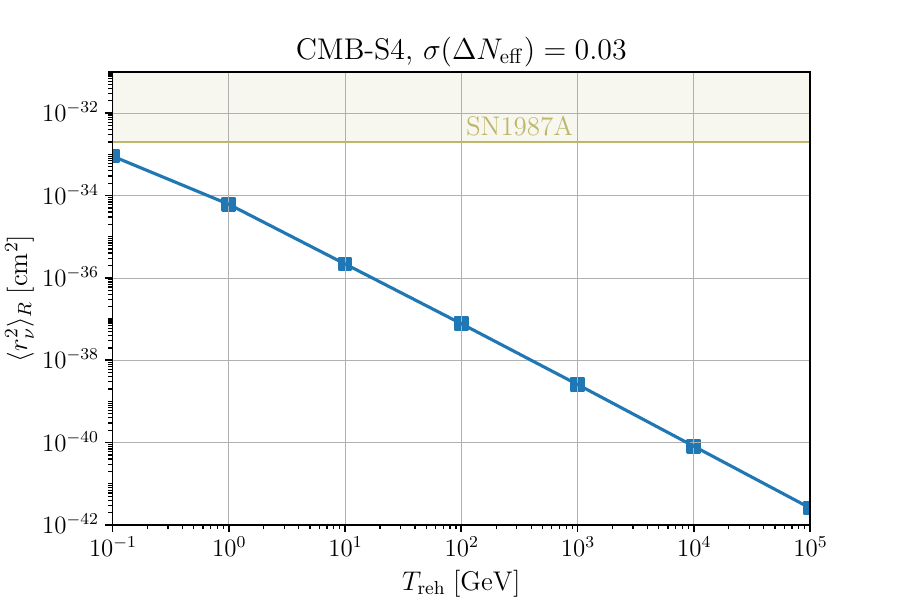}
    \caption{Forecasted $1\sigma$ upper bound on the neutrino charge radius $\langle r_\nu^2 \rangle_R$ as a function of the reheating temperature for CMB-S4, assuming $\sigma(\Delta N_{\rm eff})_{\rm{CMB-}S4} = 0.03$ \cite{CMB-S4:2016ple}. The yellow shaded region is excluded by SN1987A energy loss considerations \cite{Grifols:1989vi}.}
    \label{fig:S4-charge-radius}
\end{figure}

We can now evaluate the contribution to $\Delta N_{\rm eff}$ from right-handed neutrinos by solving the Boltzmann equation \eqref{eq:Boltzmann} ($l = 2$) with the rate \eqref{eq:rate-charge-radius} and then using Eq. \eqref{eq:DNeff}. We show our result in Fig. \ref{fig:DNeff-charge radius} for different reheating temperatures. Given the limit on $\Delta N_{\rm eff}$ derived from Planck at 95\% CL \cite{Planck:2018vyg}, we find the following bounds on the charge radius of right-handed neutrinos:
\begin{equation}
    \begin{cases}
        \langle r_\nu^2 \rangle_R < 1.87 \times 10^{-33} \; {\rm cm}^2  \quad T_{\rm reh} = 100 \; {\rm MeV} \, , \\
        \langle r_\nu^2 \rangle_R < 6.11 \times 10^{-34} \; {\rm cm}^2 \quad T_{\rm reh} \ge 1 \; {\rm GeV} \, .
    \end{cases}
\end{equation}
Note that for reheating temperatures $T_{\rm reh} \ge 1 \; {\rm GeV}$ the cosmological bound is stronger than the one derived from SN1987A. For RH$\nu$s in thermal equilibrium with the plasma and decoupling at temperatures higher than the mass of the top quark, the contribution to $\Delta N_{\rm eff}$ is slightly below the 1$\sigma$ sensitivity of SO. This agrees with the well-known result reported in Eq. \eqref{eq:DNeff-dof}. However, CMB-S4 will be sensitive to values of $\langle r_\nu^2 \rangle_R$ several orders of magnitude smaller than the bound derived from Planck, depending on the value of the reheating temperature. This is shown in Fig. \ref{fig:S4-charge-radius}.

\section{Conclusions}
\label{sec:conclusions}

Light relics are typical of BSM physics scenarios and can leave detectable imprints on cosmology. In particular, light relics often interact with the SM through non-renormalizable operators which encode the UV dynamics of BSM theories in dimensionful effective coupling constants and, thus, the corresponding interaction rates can increase with temperature more quickly than the Hubble rate. In this work, we investigate UV freeze-in production of light relics in such scenarios. UV freeze-in occurs at the highest temperature for which the relevant processes are efficient, which we assume to be the reheat temperature. Thus, while UV freeze-in is more dependent on initial conditions than freeze-out production, we demonstrate how current and next-generation CMB observations can be sensitive to coupling constants similar to or smaller than what can be measured in laboratory, astrophysical or collider probes.

We systematically explore the UV freeze-in parameter space for several representative benchmark BSM physics models. For each benchmark we demonstrate that next-generation CMB observations will be sensitive to broad regions of parameter space currently not constrained by Planck measurements of $\Delta N_{\rm eff}$. As discussed in Sec.~\ref{sec:UV-freezein}, dimensional analysis suggests that the effects of UV freeze-in are more pronounced for higher-dimensional operators mediating interactions between the light relic and the SM. In general, we see more improvement of sensitivity to UV freeze-in scenarios at next-generation experiments for benchmarks with higher dimensional operators which are relevant at higher temperatures. Thus, we see that DP production below the electroweak phase transition and ALP production are less sensitive to the temperature dependence of UV freeze-in than DP production above the electroweak phase transition and RH$\nu$ production.

For ALPs produced via the Primakoff effect at low reheat temperatures $\sim 100 \,$MeV, we show that CMB-S4 could probe ALP-photon couplings a factor of $\sim 3$ smaller than what is constrained by Planck. Futuristic proposal such as CMB-HD could probe ALP-photon couplings smaller than what is currently accessible by laboratory and astrophysical probes, $g_{a \gamma} \lesssim 6.6 \times 10^{-11} \, \mathrm{GeV}^{-1}$, for reheat temperatures $\gtrsim 10 \,$TeV. For DPs produced above the electroweak phase transition with reheat temperature up to the $\sim \,$PeV scale, we show that next-generation CMB experiments will probe effective couplings $10^{-17} \, \mathrm{GeV}^{-2} \lesssim \sqrt{\alpha}/M^2 \lesssim 10^{-12} \, \mathrm{GeV}^{-2}$ to which Planck is not at all sensitive. At low reheat temperatures $\sim 1 \,$GeV, CMB-S4 could probe freeze-in production of DPs through, for example, dipole interactions with tau leptons with effective coupling constants $\sim 2$ orders of magnitude smaller than Planck. 

For RH$\nu$s produced at reheat temperatures $\sim 1 \,$GeV through SM fermion annihilation mediated by a new gauge boson $Z'$ associated with $B-L$, CMB-S4 can probe gauge couplings a factor of $\sim 3$ smaller than Planck for gauge boson masses $M_{Z'} \gtrsim 5 \,$TeV beyond the reach of collider experiments. For $M_Z' \simeq 2 \,$TeV, CMB-S4 could probe gauge boson couplings one order of magnitude smaller than current collider constraints for reheat temperatures of $\sim 60 \,$GeV. For RH$\nu$s produced by interactions induced by the neutrino charge radius, we show that current Planck constraints are stronger than astrophysical bounds from SN1987A for reheat temperatures $\gtrsim 1 \,$GeV. For reheat temperatures $\sim 1 \,$TeV, CMB-S4 can be sensitive to $\langle r_\nu^2 \rangle_R$ around $\sim 6$ orders of magnitude smaller than astrophysical bounds.

While this work demonstrates the sensitivity of current and next-generation CMB observations to UV freeze-in of light relics for a variety of representative benchmarks, we leave more detailed analyses of particular scenarios and a more generalized approach for future studies. For example, solving the Boltzmann equation for the evolution of the light relic number density assuming thermal distribution functions and ignoring the effects of quantum statistics are approximations which can, in principle, introduce significant errors in the calculation of $\Delta N_{\rm eff}$. Recent work indicates however that such effects are likely not significant for the freeze-in of light relics produced through the scattering and annihilation processes considered in this paper~\cite{DEramo:2020gpr,Adshead:2022ovo,DEramo:2023nzt}. In any case, we plan to investigate contributions to $\Delta N_{\rm eff}$ from UV freeze-in of the light relics studied in this work using more general, momentum-dependent Boltzmann equations. 

We also plan to investigate other models for which a more detailed analysis may be necessary given the expected precision of $\Delta N_{\rm eff}$ measurements at next-generation CMB experiments. In particular, it has been shown that momentum-dependent Boltzmann equations are required to track the highly non-thermal distribution functions characteristic of light relic production through particle decays~\cite{DEramo:2023nzt}. For instance, scenarios in which Majorons, ALPs associated with the spontaneous breaking of ungauged lepton number in see-saw models~\cite{Chikashige:1980ui}, are produced in the decays of RH$\nu$s require a more detailed analysis. In addition, using momentum-dependent Boltzmann equations similar to those implemented for RH$\nu$ production induced by the neutrino magnetic moment~\cite{Carenza:2022ngg}, we plan to further investigate CMB probes of neutrino electromagnetic properties. More generally, it would be interesting to investigate how next-generation CMB observations could be sensitive to distribution functions for light relics associated with different BSM scenarios and production mechanisms.

\acknowledgments

The authors would like to thank Francesco D'Eramo for useful discussions.
PS is funded by the Istituto Nazionale di Fisica Nucleare (INFN) through the project of
the InDark INFN Special Initiative: ``Neutrinos and other light relics in view of future cosmological observations'' (n.
23590/2021). 
PS would like to thank the organizers of CATCH22+2 for their hospitality.
We acknowledge financial support from the INFN InDark initiative and from the COSMOS network (www.cosmosnet.it) through the ASI (Italian Space Agency) Grants 2016-24-H.0 and 2016-24-H.1-2018, as well as 2020-9-HH.0 (participation in LiteBIRD phase A). We acknowledge the use of numpy \cite{harris2020array} and matplotlib \cite{Hunter:2007} software packages and the use of computing facilities at CINECA. MG and ML are funded by the European Union (ERC, RELiCS, project number 101116027). MG is funded by the PRIN (Progetti di ricerca di Rilevante Interesse Nazionale) number 2022WJ9J33. This article is based upon work from COST Action COSMIC WISPers CA21106, supported by COST (European Cooperation in Science and Technology).

\bibliographystyle{JHEP}
\bibliography{bibliography.bib}

\providecommand{\href}[2]{#2}\begingroup\raggedright\begin{thebibliography}{10}

\bibitem{PhysRevLett.40.223}
S.~Weinberg, \emph{A new light boson?}, \href{http://dx.doi.org/10.1103/PhysRevLett.40.223}{\emph{Phys. Rev. Lett.} {\bf 40} (Jan, 1978) 223--226}.

\bibitem{PhysRevLett.40.279}
F.~Wilczek, \emph{Problem of strong $p$ and $t$ invariance in the presence of instantons}, \href{http://dx.doi.org/10.1103/PhysRevLett.40.279}{\emph{Phys. Rev. Lett.} {\bf 40} (Jan, 1978) 279--282}.

\bibitem{PhysRevLett.40.220}
T.~Goldman and C.~M. Hoffman, \emph{Will the axion be found soon?}, \href{http://dx.doi.org/10.1103/PhysRevLett.40.220}{\emph{Phys. Rev. Lett.} {\bf 40} (Jan, 1978) 220--222}.

\bibitem{McDonald:1993ex}
J.~McDonald, \emph{{Gauge singlet scalars as cold dark matter}}, \href{http://dx.doi.org/10.1103/PhysRevD.50.3637}{\emph{Phys. Rev. D} {\bf 50} (1994) 3637--3649}, [\href{http://arxiv.org/abs/hep-ph/0702143}{{\tt hep-ph/0702143}}].

\bibitem{Burgess:2000yq}
C.~P. Burgess, M.~Pospelov and T.~ter Veldhuis, \emph{{The Minimal model of nonbaryonic dark matter: A Singlet scalar}}, \href{http://dx.doi.org/10.1016/S0550-3213(01)00513-2}{\emph{Nucl. Phys. B} {\bf 619} (2001) 709--728}, [\href{http://arxiv.org/abs/hep-ph/0011335}{{\tt hep-ph/0011335}}].

\bibitem{He:2009yd}
X.-G. He, T.~Li, X.-Q. Li, J.~Tandean and H.-C. Tsai, \emph{{The Simplest Dark-Matter Model, CDMS II Results, and Higgs Detection at LHC}}, \href{http://dx.doi.org/10.1016/j.physletb.2010.04.026}{\emph{Phys. Lett. B} {\bf 688} (2010) 332--336}, [\href{http://arxiv.org/abs/0912.4722}{{\tt 0912.4722}}].

\bibitem{Cacciapaglia:2019bqz}
G.~Cacciapaglia, G.~Ferretti, T.~Flacke and H.~Ser\^odio, \emph{{Light scalars in composite Higgs models}}, \href{http://dx.doi.org/10.3389/fphy.2019.00022}{\emph{Front. in Phys.} {\bf 7} (2019) 22}, [\href{http://arxiv.org/abs/1902.06890}{{\tt 1902.06890}}].

\bibitem{Svrcek:2006yi}
P.~Svrcek and E.~Witten, \emph{{Axions In String Theory}}, \href{http://dx.doi.org/10.1088/1126-6708/2006/06/051}{\emph{JHEP} {\bf 06} (2006) 051}, [\href{http://arxiv.org/abs/hep-th/0605206}{{\tt hep-th/0605206}}].

\bibitem{Arvanitaki:2009fg}
A.~Arvanitaki, S.~Dimopoulos, S.~Dubovsky, N.~Kaloper and J.~March-Russell, \emph{{String Axiverse}}, \href{http://dx.doi.org/10.1103/PhysRevD.81.123530}{\emph{Phys. Rev. D} {\bf 81} (2010) 123530}, [\href{http://arxiv.org/abs/0905.4720}{{\tt 0905.4720}}].

\bibitem{Holdom:1985ag}
B.~Holdom, \emph{{Two U(1)'s and Epsilon Charge Shifts}}, \href{http://dx.doi.org/10.1016/0370-2693(86)91377-8}{\emph{Phys. Lett. B} {\bf 166} (1986) 196--198}.

\bibitem{Fabbrichesi:2020wbt}
M.~Fabbrichesi, E.~Gabrielli and G.~Lanfranchi, \emph{{The Dark Photon}},  \href{http://arxiv.org/abs/2005.01515}{{\tt 2005.01515}}.

\bibitem{Abel:2003ue}
S.~A. Abel and B.~W. Schofield, \emph{{Brane anti-brane kinetic mixing, millicharged particles and SUSY breaking}}, \href{http://dx.doi.org/10.1016/j.nuclphysb.2004.02.037}{\emph{Nucl. Phys. B} {\bf 685} (2004) 150--170}, [\href{http://arxiv.org/abs/hep-th/0311051}{{\tt hep-th/0311051}}].

\bibitem{Abel:2006qt}
S.~A. Abel, J.~Jaeckel, V.~V. Khoze and A.~Ringwald, \emph{{Illuminating the Hidden Sector of String Theory by Shining Light through a Magnetic Field}}, \href{http://dx.doi.org/10.1016/j.physletb.2008.03.076}{\emph{Phys. Lett. B} {\bf 666} (2008) 66--70}, [\href{http://arxiv.org/abs/hep-ph/0608248}{{\tt hep-ph/0608248}}].

\bibitem{Abel:2008ai}
S.~A. Abel, M.~D. Goodsell, J.~Jaeckel, V.~V. Khoze and A.~Ringwald, \emph{{Kinetic Mixing of the Photon with Hidden U(1)s in String Phenomenology}}, \href{http://dx.doi.org/10.1088/1126-6708/2008/07/124}{\emph{JHEP} {\bf 07} (2008) 124}, [\href{http://arxiv.org/abs/0803.1449}{{\tt 0803.1449}}].

\bibitem{Goodsell:2009xc}
M.~Goodsell, J.~Jaeckel, J.~Redondo and A.~Ringwald, \emph{{Naturally Light Hidden Photons in LARGE Volume String Compactifications}}, \href{http://dx.doi.org/10.1088/1126-6708/2009/11/027}{\emph{JHEP} {\bf 11} (2009) 027}, [\href{http://arxiv.org/abs/0909.0515}{{\tt 0909.0515}}].

\bibitem{Berezhiani:2003xm}
Z.~Berezhiani, \emph{{Mirror world and its cosmological consequences}}, \href{http://dx.doi.org/10.1142/S0217751X04020075}{\emph{Int. J. Mod. Phys. A} {\bf 19} (2004) 3775--3806}, [\href{http://arxiv.org/abs/hep-ph/0312335}{{\tt hep-ph/0312335}}].

\bibitem{Berezhiani:2008gi}
Z.~Berezhiani and A.~Lepidi, \emph{{Cosmological bounds on the 'millicharges' of mirror particles}}, \href{http://dx.doi.org/10.1016/j.physletb.2009.10.023}{\emph{Phys. Lett. B} {\bf 681} (2009) 276--281}, [\href{http://arxiv.org/abs/0810.1317}{{\tt 0810.1317}}].

\bibitem{Salvio:2014soa}
A.~Salvio and A.~Strumia, \emph{{Agravity}}, \href{http://dx.doi.org/10.1007/JHEP06(2014)080}{\emph{JHEP} {\bf 06} (2014) 080}, [\href{http://arxiv.org/abs/1403.4226}{{\tt 1403.4226}}].

\bibitem{Salvio:2020prd}
A.~Salvio, \emph{{A fundamental QCD axion model}}, \href{http://dx.doi.org/10.1016/j.physletb.2020.135686}{\emph{Phys. Lett. B} {\bf 808} (2020) 135686}, [\href{http://arxiv.org/abs/2003.10446}{{\tt 2003.10446}}].

\bibitem{Ghoshal:2020vud}
A.~Ghoshal and A.~Salvio, \emph{{Gravitational waves from fundamental axion dynamics}}, \href{http://dx.doi.org/10.1007/JHEP12(2020)049}{\emph{JHEP} {\bf 12} (2020) 049}, [\href{http://arxiv.org/abs/2007.00005}{{\tt 2007.00005}}].

\bibitem{Mohapatra:2005wg}
R.~N. Mohapatra et~al., \emph{{Theory of neutrinos: A White paper}}, \href{http://dx.doi.org/10.1088/0034-4885/70/11/R02}{\emph{Rept. Prog. Phys.} {\bf 70} (2007) 1757--1867}, [\href{http://arxiv.org/abs/hep-ph/0510213}{{\tt hep-ph/0510213}}].

\bibitem{King:2013eh}
S.~F. King and C.~Luhn, \emph{{Neutrino Mass and Mixing with Discrete Symmetry}}, \href{http://dx.doi.org/10.1088/0034-4885/76/5/056201}{\emph{Rept. Prog. Phys.} {\bf 76} (2013) 056201}, [\href{http://arxiv.org/abs/1301.1340}{{\tt 1301.1340}}].

\bibitem{King:2015aea}
S.~F. King, \emph{{Models of Neutrino Mass, Mixing and CP Violation}}, \href{http://dx.doi.org/10.1088/0954-3899/42/12/123001}{\emph{J. Phys. G} {\bf 42} (2015) 123001}, [\href{http://arxiv.org/abs/1510.02091}{{\tt 1510.02091}}].

\bibitem{Carlson:1986cu}
E.~D. Carlson, \emph{{LIMITS ON A NEW U(1) COUPLING}}, \href{http://dx.doi.org/10.1016/0550-3213(87)90446-9}{\emph{Nucl. Phys. B} {\bf 286} (1987) 378--398}.

\bibitem{Feldman:2011ms}
D.~Feldman, P.~Fileviez~Perez and P.~Nath, \emph{{R-parity Conservation via the Stueckelberg Mechanism: LHC and Dark Matter Signals}}, \href{http://dx.doi.org/10.1007/JHEP01(2012)038}{\emph{JHEP} {\bf 01} (2012) 038}, [\href{http://arxiv.org/abs/1109.2901}{{\tt 1109.2901}}].

\bibitem{Dvorkin:2022jyg}
C.~Dvorkin et~al., \emph{{The Physics of Light Relics}},  in \emph{{Snowmass 2021}}, 3, 2022.
\newblock \href{http://arxiv.org/abs/2203.07943}{{\tt 2203.07943}}.

\bibitem{Bashinsky:2003tk}
S.~Bashinsky and U.~Seljak, \emph{{Neutrino perturbations in CMB anisotropy and matter clustering}}, \href{http://dx.doi.org/10.1103/PhysRevD.69.083002}{\emph{Phys. Rev. D} {\bf 69} (2004) 083002}, [\href{http://arxiv.org/abs/astro-ph/0310198}{{\tt astro-ph/0310198}}].

\bibitem{Hou:2011ec}
Z.~Hou, R.~Keisler, L.~Knox, M.~Millea and C.~Reichardt, \emph{{How Massless Neutrinos Affect the Cosmic Microwave Background Damping Tail}}, \href{http://dx.doi.org/10.1103/PhysRevD.87.083008}{\emph{Phys. Rev. D} {\bf 87} (2013) 083008}, [\href{http://arxiv.org/abs/1104.2333}{{\tt 1104.2333}}].

\bibitem{Baumann:2015rya}
D.~Baumann, D.~Green, J.~Meyers and B.~Wallisch, \emph{{Phases of New Physics in the CMB}}, \href{http://dx.doi.org/10.1088/1475-7516/2016/01/007}{\emph{JCAP} {\bf 01} (2016) 007}, [\href{http://arxiv.org/abs/1508.06342}{{\tt 1508.06342}}].

\bibitem{Brust:2013ova}
C.~Brust, D.~E. Kaplan and M.~T. Walters, \emph{{New Light Species and the CMB}}, \href{http://dx.doi.org/10.1007/JHEP12(2013)058}{\emph{JHEP} {\bf 12} (2013) 058}, [\href{http://arxiv.org/abs/1303.5379}{{\tt 1303.5379}}].

\bibitem{Chacko:2015noa}
Z.~Chacko, Y.~Cui, S.~Hong and T.~Okui, \emph{{Hidden dark matter sector, dark radiation, and the CMB}}, \href{http://dx.doi.org/10.1103/PhysRevD.92.055033}{\emph{Phys. Rev. D} {\bf 92} (2015) 055033}, [\href{http://arxiv.org/abs/1505.04192}{{\tt 1505.04192}}].

\bibitem{Baumann:2016wac}
D.~Baumann, D.~Green and B.~Wallisch, \emph{{New Target for Cosmic Axion Searches}}, \href{http://dx.doi.org/10.1103/PhysRevLett.117.171301}{\emph{Phys. Rev. Lett.} {\bf 117} (2016) 171301}, [\href{http://arxiv.org/abs/1604.08614}{{\tt 1604.08614}}].

\bibitem{Hall:2009bx}
L.~J. Hall, K.~Jedamzik, J.~March-Russell and S.~M. West, \emph{{Freeze-In Production of FIMP Dark Matter}}, \href{http://dx.doi.org/10.1007/JHEP03(2010)080}{\emph{JHEP} {\bf 03} (2010) 080}, [\href{http://arxiv.org/abs/0911.1120}{{\tt 0911.1120}}].

\bibitem{Cheung:2010gj}
C.~Cheung, G.~Elor, L.~J. Hall and P.~Kumar, \emph{{Origins of Hidden Sector Dark Matter I: Cosmology}}, \href{http://dx.doi.org/10.1007/JHEP03(2011)042}{\emph{JHEP} {\bf 03} (2011) 042}, [\href{http://arxiv.org/abs/1010.0022}{{\tt 1010.0022}}].

\bibitem{Hall:2010jx}
L.~J. Hall, J.~March-Russell and S.~M. West, \emph{{A Unified Theory of Matter Genesis: Asymmetric Freeze-In}},  \href{http://arxiv.org/abs/1010.0245}{{\tt 1010.0245}}.

\bibitem{Elahi:2014fsa}
F.~Elahi, C.~Kolda and J.~Unwin, \emph{{UltraViolet Freeze-in}}, \href{http://dx.doi.org/10.1007/JHEP03(2015)048}{\emph{JHEP} {\bf 03} (2015) 048}, [\href{http://arxiv.org/abs/1410.6157}{{\tt 1410.6157}}].

\bibitem{DEramo:2020gpr}
F.~D'Eramo and A.~Lenoci, \emph{{Lower mass bounds on FIMP dark matter produced via freeze-in}}, \href{http://dx.doi.org/10.1088/1475-7516/2021/10/045}{\emph{JCAP} {\bf 10} (2021) 045}, [\href{http://arxiv.org/abs/2012.01446}{{\tt 2012.01446}}].

\bibitem{Mangano:2001iu}
G.~Mangano, G.~Miele, S.~Pastor and M.~Peloso, \emph{{A Precision calculation of the effective number of cosmological neutrinos}}, \href{http://dx.doi.org/10.1016/S0370-2693(02)01622-2}{\emph{Phys. Lett. B} {\bf 534} (2002) 8--16}, [\href{http://arxiv.org/abs/astro-ph/0111408}{{\tt astro-ph/0111408}}].

\bibitem{Bennett:2019ewm}
J.~J. Bennett, G.~Buldgen, M.~Drewes and Y.~Y.~Y. Wong, \emph{{Towards a precision calculation of the effective number of neutrinos $N_{\rm eff}$ in the Standard Model I: the QED equation of state}}, \href{http://dx.doi.org/10.1088/1475-7516/2020/03/003}{\emph{JCAP} {\bf 03} (2020) 003}, [\href{http://arxiv.org/abs/1911.04504}{{\tt 1911.04504}}].

\bibitem{Bennett:2020zkv}
J.~J. Bennett, G.~Buldgen, P.~F. De~Salas, M.~Drewes, S.~Gariazzo, S.~Pastor et~al., \emph{{Towards a precision calculation of $N_{\rm eff}$ in the Standard Model II: Neutrino decoupling in the presence of flavour oscillations and finite-temperature QED}}, \href{http://dx.doi.org/10.1088/1475-7516/2021/04/073}{\emph{JCAP} {\bf 04} (2021) 073}, [\href{http://arxiv.org/abs/2012.02726}{{\tt 2012.02726}}].

\bibitem{Akita:2020szl}
K.~Akita and M.~Yamaguchi, \emph{{A precision calculation of relic neutrino decoupling}}, \href{http://dx.doi.org/10.1088/1475-7516/2020/08/012}{\emph{JCAP} {\bf 08} (2020) 012}, [\href{http://arxiv.org/abs/2005.07047}{{\tt 2005.07047}}].

\bibitem{Froustey:2020mcq}
J.~Froustey, C.~Pitrou and M.~C. Volpe, \emph{{Neutrino decoupling including flavour oscillations and primordial nucleosynthesis}}, \href{http://dx.doi.org/10.1088/1475-7516/2020/12/015}{\emph{JCAP} {\bf 12} (2020) 015}, [\href{http://arxiv.org/abs/2008.01074}{{\tt 2008.01074}}].

\bibitem{Cielo:2023bqp}
M.~Cielo, M.~Escudero, G.~Mangano and O.~Pisanti, \emph{{Neff in the Standard Model at NLO is 3.043}}, \href{http://dx.doi.org/10.1103/PhysRevD.108.L121301}{\emph{Phys. Rev. D} {\bf 108} (2023) L121301}, [\href{http://arxiv.org/abs/2306.05460}{{\tt 2306.05460}}].

\bibitem{Drewes:2024wbw}
M.~Drewes, Y.~Georis, M.~Klasen, L.~P. Wiggering and Y.~Y.~Y. Wong, \emph{{Towards a precision calculation of $N_{\rm eff}$ in the Standard Model III: Improved estimate of NLO corrections to the collision integral}},  \href{http://arxiv.org/abs/2402.18481}{{\tt 2402.18481}}.

\bibitem{Planck:2018vyg}
{\scshape Planck} collaboration, N.~Aghanim et~al., \emph{{Planck 2018 results. VI. Cosmological parameters}}, \href{http://dx.doi.org/10.1051/0004-6361/201833910}{\emph{Astron. Astrophys.} {\bf 641} (2020) A6}, [\href{http://arxiv.org/abs/1807.06209}{{\tt 1807.06209}}].

\bibitem{SimonsObservatory:2018koc}
{\scshape Simons Observatory} collaboration, P.~Ade et~al., \emph{{The Simons Observatory: Science goals and forecasts}}, \href{http://dx.doi.org/10.1088/1475-7516/2019/02/056}{\emph{JCAP} {\bf 02} (2019) 056}, [\href{http://arxiv.org/abs/1808.07445}{{\tt 1808.07445}}].

\bibitem{CMB-S4:2016ple}
{\scshape CMB-S4} collaboration, K.~N. Abazajian et~al., \emph{{CMB-S4 Science Book, First Edition}},  \href{http://arxiv.org/abs/1610.02743}{{\tt 1610.02743}}.

\bibitem{CMB-HD:2022bsz}
{\scshape CMB-HD} collaboration, S.~Aiola et~al., \emph{{Snowmass2021 CMB-HD White Paper}},  \href{http://arxiv.org/abs/2203.05728}{{\tt 2203.05728}}.

\bibitem{DEramo:2022nvb}
F.~D'Eramo, E.~Di~Valentino, W.~Giar\`e, F.~Hajkarim, A.~Melchiorri, O.~Mena et~al., \emph{{Cosmological bound on the QCD axion mass, redux}}, \href{http://dx.doi.org/10.1088/1475-7516/2022/09/022}{\emph{JCAP} {\bf 09} (2022) 022}, [\href{http://arxiv.org/abs/2205.07849}{{\tt 2205.07849}}].

\bibitem{Notari:2022ffe}
A.~Notari, F.~Rompineve and G.~Villadoro, \emph{{Improved Hot Dark Matter Bound on the QCD Axion}}, \href{http://dx.doi.org/10.1103/PhysRevLett.131.011004}{\emph{Phys. Rev. Lett.} {\bf 131} (2023) 011004}, [\href{http://arxiv.org/abs/2211.03799}{{\tt 2211.03799}}].

\bibitem{Bianchini:2023ubu}
F.~Bianchini, G.~G. di~Cortona and M.~Valli, \emph{{The QCD Axion: Some Like It Hot}},  \href{http://arxiv.org/abs/2310.08169}{{\tt 2310.08169}}.

\bibitem{Dror:2021nyr}
J.~A. Dror, H.~Murayama and N.~L. Rodd, \emph{{Cosmic axion background}}, \href{http://dx.doi.org/10.1103/PhysRevD.103.115004}{\emph{Phys. Rev. D} {\bf 103} (2021) 115004}, [\href{http://arxiv.org/abs/2101.09287}{{\tt 2101.09287}}].

\bibitem{Caloni:2022uya}
L.~Caloni, M.~Gerbino, M.~Lattanzi and L.~Visinelli, \emph{{Novel cosmological bounds on thermally-produced axion-like particles}}, \href{http://dx.doi.org/10.1088/1475-7516/2022/09/021}{\emph{JCAP} {\bf 09} (2022) 021}, [\href{http://arxiv.org/abs/2205.01637}{{\tt 2205.01637}}].

\bibitem{DEramo:2023nzt}
F.~D'Eramo, F.~Hajkarim and A.~Lenoci, \emph{{Dark radiation from the primordial thermal bath in momentum space}}, \href{http://dx.doi.org/10.1088/1475-7516/2024/03/009}{\emph{JCAP} {\bf 03} (2024) 009}, [\href{http://arxiv.org/abs/2311.04974}{{\tt 2311.04974}}].

\bibitem{Bolz:2000fu}
M.~Bolz, A.~Brandenburg and W.~Buchmuller, \emph{{Thermal production of gravitinos}}, \href{http://dx.doi.org/10.1016/S0550-3213(01)00132-8}{\emph{Nucl. Phys. B} {\bf 606} (2001) 518--544}, [\href{http://arxiv.org/abs/hep-ph/0012052}{{\tt hep-ph/0012052}}].

\bibitem{Cadamuro:2011fd}
D.~Cadamuro and J.~Redondo, \emph{{Cosmological bounds on pseudo Nambu-Goldstone bosons}}, \href{http://dx.doi.org/10.1088/1475-7516/2012/02/032}{\emph{JCAP} {\bf 02} (2012) 032}, [\href{http://arxiv.org/abs/1110.2895}{{\tt 1110.2895}}].

\bibitem{Dobrescu:2004wz}
B.~A. Dobrescu, \emph{{Massless gauge bosons other than the photon}}, \href{http://dx.doi.org/10.1103/PhysRevLett.94.151802}{\emph{Phys. Rev. Lett.} {\bf 94} (2005) 151802}, [\href{http://arxiv.org/abs/hep-ph/0411004}{{\tt hep-ph/0411004}}].

\bibitem{Vogel:2013raa}
H.~Vogel and J.~Redondo, \emph{{Dark Radiation constraints on minicharged particles in models with a hidden photon}}, \href{http://dx.doi.org/10.1088/1475-7516/2014/02/029}{\emph{JCAP} {\bf 02} (2014) 029}, [\href{http://arxiv.org/abs/1311.2600}{{\tt 1311.2600}}].

\bibitem{Foot:2014uba}
R.~Foot and S.~Vagnozzi, \emph{{Dissipative hidden sector dark matter}}, \href{http://dx.doi.org/10.1103/PhysRevD.91.023512}{\emph{Phys. Rev. D} {\bf 91} (2015) 023512}, [\href{http://arxiv.org/abs/1409.7174}{{\tt 1409.7174}}].

\bibitem{Salvio:2022hfa}
A.~Salvio, \emph{{Thermal production of massless dark photons}}, \href{http://dx.doi.org/10.1088/1475-7516/2023/07/035}{\emph{JCAP} {\bf 07} (2023) 035}, [\href{http://arxiv.org/abs/2212.09755}{{\tt 2212.09755}}].

\bibitem{Adshead:2022ovo}
P.~Adshead, P.~Ralegankar and J.~Shelton, \emph{{Dark radiation constraints on portal interactions with hidden sectors}}, \href{http://dx.doi.org/10.1088/1475-7516/2022/09/056}{\emph{JCAP} {\bf 09} (2022) 056}, [\href{http://arxiv.org/abs/2206.13530}{{\tt 2206.13530}}].

\bibitem{Barger:2003zh}
V.~Barger, P.~Langacker and H.-S. Lee, \emph{{Primordial nucleosynthesis constraints on $Z^\prime$ properties}}, \href{http://dx.doi.org/10.1103/PhysRevD.67.075009}{\emph{Phys. Rev. D} {\bf 67} (2003) 075009}, [\href{http://arxiv.org/abs/hep-ph/0302066}{{\tt hep-ph/0302066}}].

\bibitem{Heeck:2014zfa}
J.~Heeck, \emph{{Unbroken B \textendash{} L symmetry}}, \href{http://dx.doi.org/10.1016/j.physletb.2014.10.067}{\emph{Phys. Lett. B} {\bf 739} (2014) 256--262}, [\href{http://arxiv.org/abs/1408.6845}{{\tt 1408.6845}}].

\bibitem{FileviezPerez:2019cyn}
P.~Fileviez~P\'erez, C.~Murgui and A.~D. Plascencia, \emph{{Neutrino-Dark Matter Connections in Gauge Theories}}, \href{http://dx.doi.org/10.1103/PhysRevD.100.035041}{\emph{Phys. Rev. D} {\bf 100} (2019) 035041}, [\href{http://arxiv.org/abs/1905.06344}{{\tt 1905.06344}}].

\bibitem{Abazajian:2019oqj}
K.~N. Abazajian and J.~Heeck, \emph{{Observing Dirac neutrinos in the cosmic microwave background}}, \href{http://dx.doi.org/10.1103/PhysRevD.100.075027}{\emph{Phys. Rev. D} {\bf 100} (2019) 075027}, [\href{http://arxiv.org/abs/1908.03286}{{\tt 1908.03286}}].

\bibitem{Bernabeu:2004jr}
J.~Bernabeu, J.~Papavassiliou and D.~Binosi, \emph{{The Neutrino charge radius in the presence of fermion masses}}, \href{http://dx.doi.org/10.1016/j.nuclphysb.2005.02.039}{\emph{Nucl. Phys. B} {\bf 716} (2005) 352--372}, [\href{http://arxiv.org/abs/hep-ph/0405288}{{\tt hep-ph/0405288}}].

\bibitem{Elmfors:1997tt}
P.~Elmfors, K.~Enqvist, G.~Raffelt and G.~Sigl, \emph{{Neutrinos with magnetic moment: Depolarization rate in plasma}}, \href{http://dx.doi.org/10.1016/S0550-3213(97)00382-9}{\emph{Nucl. Phys. B} {\bf 503} (1997) 3--23}, [\href{http://arxiv.org/abs/hep-ph/9703214}{{\tt hep-ph/9703214}}].

\bibitem{Li:2022dkc}
S.-P. Li and X.-J. Xu, \emph{{Neutrino magnetic moments meet precision N$_{eff}$ measurements}}, \href{http://dx.doi.org/10.1007/JHEP02(2023)085}{\emph{JHEP} {\bf 02} (2023) 085}, [\href{http://arxiv.org/abs/2211.04669}{{\tt 2211.04669}}].

\bibitem{Carenza:2022ngg}
P.~Carenza, G.~Lucente, M.~Gerbino, M.~Giannotti and M.~Lattanzi, \emph{{Strong cosmological constraints on the neutrino magnetic moment}},  \href{http://arxiv.org/abs/2211.10432}{{\tt 2211.10432}}.

\bibitem{Grifols:1986ed}
J.~A. Grifols and E.~Masso, \emph{{Bound on the Neutrino Charge Radius From Primordial Nucleosynthesis}}, \href{http://dx.doi.org/10.1142/S0217732387000276}{\emph{Mod. Phys. Lett. A} {\bf 2} (1987) 205}.

\bibitem{Grifols:1989vi}
J.~A. Grifols and E.~Masso, \emph{{Charge Radius of the Neutrino: A Limit From {SN1987A}}}, \href{http://dx.doi.org/10.1103/PhysRevD.40.3819}{\emph{Phys. Rev. D} {\bf 40} (1989) 3819}.

\bibitem{Saikawa:2018rcs}
K.~Saikawa and S.~Shirai, \emph{{Primordial gravitational waves, precisely: The role of thermodynamics in the Standard Model}}, \href{http://dx.doi.org/10.1088/1475-7516/2018/05/035}{\emph{JCAP} {\bf 05} (2018) 035}, [\href{http://arxiv.org/abs/1803.01038}{{\tt 1803.01038}}].

\bibitem{DEramo:2021lgb}
F.~D'Eramo, F.~Hajkarim and S.~Yun, \emph{{Thermal QCD Axions across Thresholds}}, \href{http://dx.doi.org/10.1007/JHEP10(2021)224}{\emph{JHEP} {\bf 10} (2021) 224}, [\href{http://arxiv.org/abs/2108.05371}{{\tt 2108.05371}}].

\bibitem{CAST:2017uph}
{\scshape CAST} collaboration, V.~Anastassopoulos et~al., \emph{{New CAST Limit on the Axion-Photon Interaction}}, \href{http://dx.doi.org/10.1038/nphys4109}{\emph{Nature Phys.} {\bf 13} (2017) 584--590}, [\href{http://arxiv.org/abs/1705.02290}{{\tt 1705.02290}}].

\bibitem{Friedland:2012hj}
A.~Friedland, M.~Giannotti and M.~Wise, \emph{{Constraining the Axion-Photon Coupling with Massive Stars}}, \href{http://dx.doi.org/10.1103/PhysRevLett.110.061101}{\emph{Phys. Rev. Lett.} {\bf 110} (2013) 061101}, [\href{http://arxiv.org/abs/1210.1271}{{\tt 1210.1271}}].

\bibitem{Ayala:2014pea}
A.~Ayala, I.~Dom\'\i{}nguez, M.~Giannotti, A.~Mirizzi and O.~Straniero, \emph{{Revisiting the bound on axion-photon coupling from Globular Clusters}}, \href{http://dx.doi.org/10.1103/PhysRevLett.113.191302}{\emph{Phys. Rev. Lett.} {\bf 113} (2014) 191302}, [\href{http://arxiv.org/abs/1406.6053}{{\tt 1406.6053}}].

\bibitem{CMS:2018ktx}
{\scshape CMS} collaboration, A.~M. Sirunyan et~al., \emph{{Measurement of the weak mixing angle using the forward-backward asymmetry of Drell-Yan events in pp collisions at 8 TeV}}, \href{http://dx.doi.org/10.1140/epjc/s10052-018-6148-7}{\emph{Eur. Phys. J. C} {\bf 78} (2018) 701}, [\href{http://arxiv.org/abs/1806.00863}{{\tt 1806.00863}}].

\bibitem{Gondolo:1990dk}
P.~Gondolo and G.~Gelmini, \emph{{Cosmic abundances of stable particles: Improved analysis}}, \href{http://dx.doi.org/10.1016/0550-3213(91)90438-4}{\emph{Nucl. Phys. B} {\bf 360} (1991) 145--179}.

\bibitem{Langacker:1991pg}
P.~Langacker and M.-x. Luo, \emph{{Constraints on additional $Z$ bosons}}, \href{http://dx.doi.org/10.1103/PhysRevD.45.278}{\emph{Phys. Rev. D} {\bf 45} (1992) 278--292}.

\bibitem{Workman:2022ynf}
{\scshape Particle Data Group} collaboration, R.~L. Workman and Others, \emph{{Review of Particle Physics}}, \href{http://dx.doi.org/10.1093/ptep/ptac097}{\emph{PTEP} {\bf 2022} (2022) 083C01}.

\bibitem{ATLAS:2017fih}
{\scshape ATLAS} collaboration, M.~Aaboud et~al., \emph{{Search for new high-mass phenomena in the dilepton final state using 36 fb${^{-1}}$ of proton-proton collision data at $ {\sqrt{s}=13} $ TeV with the ATLAS detector}}, \href{http://dx.doi.org/10.1007/JHEP10(2017)182}{\emph{JHEP} {\bf 10} (2017) 182}, [\href{http://arxiv.org/abs/1707.02424}{{\tt 1707.02424}}].

\bibitem{ATLAS:2017rue}
{\scshape ATLAS} collaboration, M.~Aaboud et~al., \emph{{Measurement of the Drell-Yan triple-differential cross section in $pp$ collisions at $\sqrt{s} = 8$ TeV}}, \href{http://dx.doi.org/10.1007/JHEP12(2017)059}{\emph{JHEP} {\bf 12} (2017) 059}, [\href{http://arxiv.org/abs/1710.05167}{{\tt 1710.05167}}].

\bibitem{Escudero:2018fwn}
M.~Escudero, S.~J. Witte and N.~Rius, \emph{{The dispirited case of gauged U(1)$_{B-L}$ dark matter}}, \href{http://dx.doi.org/10.1007/JHEP08(2018)190}{\emph{JHEP} {\bf 08} (2018) 190}, [\href{http://arxiv.org/abs/1806.02823}{{\tt 1806.02823}}].

\bibitem{LEP:2004xhf}
{\scshape LEP, ALEPH, DELPHI, L3, LEP Electroweak Working Group, SLD Electroweak Group, SLD Heavy Flavour Group, OPAL} collaboration, \emph{{A Combination of preliminary electroweak measurements and constraints on the standard model}},  \href{http://arxiv.org/abs/hep-ex/0412015}{{\tt hep-ex/0412015}}.

\bibitem{Appelquist:2002mw}
T.~Appelquist, B.~A. Dobrescu and A.~R. Hopper, \emph{{Nonexotic Neutral Gauge Bosons}}, \href{http://dx.doi.org/10.1103/PhysRevD.68.035012}{\emph{Phys. Rev. D} {\bf 68} (2003) 035012}, [\href{http://arxiv.org/abs/hep-ph/0212073}{{\tt hep-ph/0212073}}].

\bibitem{Carena:2004xs}
M.~Carena, A.~Daleo, B.~A. Dobrescu and T.~M.~P. Tait, \emph{{$Z^\prime$ gauge bosons at the Tevatron}}, \href{http://dx.doi.org/10.1103/PhysRevD.70.093009}{\emph{Phys. Rev. D} {\bf 70} (2004) 093009}, [\href{http://arxiv.org/abs/hep-ph/0408098}{{\tt hep-ph/0408098}}].

\bibitem{Giunti:2014ixa}
C.~Giunti and A.~Studenikin, \emph{{Neutrino electromagnetic interactions: a window to new physics}}, \href{http://dx.doi.org/10.1103/RevModPhys.87.531}{\emph{Rev. Mod. Phys.} {\bf 87} (2015) 531}, [\href{http://arxiv.org/abs/1403.6344}{{\tt 1403.6344}}].

\bibitem{Chikashige:1980ui}
Y.~Chikashige, R.~N. Mohapatra and R.~D. Peccei, \emph{{Are There Real Goldstone Bosons Associated with Broken Lepton Number?}}, \href{http://dx.doi.org/10.1016/0370-2693(81)90011-3}{\emph{Phys. Lett. B} {\bf 98} (1981) 265--268}.

\bibitem{harris2020array}
C.~R. Harris, K.~J. Millman, S.~J. van~der Walt, R.~Gommers, P.~Virtanen, D.~Cournapeau et~al., \emph{Array programming with {NumPy}}, \href{http://dx.doi.org/10.1038/s41586-020-2649-2}{\emph{Nature} {\bf 585} (Sept., 2020) 357--362}.

\bibitem{Hunter:2007}
J.~D. Hunter, \emph{Matplotlib: A 2d graphics environment}, \href{http://dx.doi.org/10.1109/MCSE.2007.55}{\emph{Computing in Science \& Engineering} {\bf 9} (2007) 90--95}.

\end{thebibliography}\endgroup

\end{document}